\documentstyle[12pt,fleqn]{article}

\font\twlgot =eufm10 scaled \magstep1
\font\egtgot =eufm8
\font\sevgot =eufm7

\font\twlmsb =msbm10 scaled \magstep1
\font\egtmsb =msbm8
\font\sevmsb =msbm7

\newfam\gotfam
\def\pgot{\fam\gotfam\twlgot}
\textfont\gotfam\twlgot
\scriptfont\gotfam\egtgot
\scriptscriptfont\gotfam\sevgot
\def\got{\protect\pgot}

\newfam\msbfam
\textfont\msbfam\twlmsb
\scriptfont\msbfam\egtmsb
\scriptscriptfont\msbfam\sevmsb
\def\Bbb{\protect\pBbb}
\def\pBbb{\relax\ifmmode\expandafter\Bb\else\typeout{You cann't use
Bbb in text mode}\fi}
\def\Bb #1{{\fam\msbfam\relax#1}}

\let\Large=\large

\def\op#1{\mathop{{\it\fam0} #1}\limits}
\newcommand{\glos}[1]{\bigskip{\bf #1}\bigskip}
\newcommand{\id}{{\rm Id\,}}

\newcommand{\pr}{{\rm pr}}
\newcommand{\df}{{\rm def}}

\newcommand{\Ker}{{\rm Ker\,}}
\newcommand{\im}{{\rm Im\, }}

\newcommand{\nm}[1]{\mid {#1}\mid}

\newcommand{\beq}{\begin{equation}}
\newcommand{\eeq}{\end{equation}}
\newcommand{\ben}{\begin{eqnarray}}
\newcommand{\een}{\end{eqnarray}}
\newcommand{\be}{\begin{eqnarray*}}
\newcommand{\ee}{\end{eqnarray*}}
\newcommand{\bea}{\begin{eqalph}}
\newcommand{\eea}{\end{eqalph}}

\newcommand{\cO}{{\got O}}

\newcommand{\cT}{{\cal T}}

\newcommand{\bL}{{\bf L}}

\newcommand{\rL}{{\rm L}}

\newcommand{\rs}{{\rm s}}

\newcommand{\al}{\alpha}
\newcommand{\bt}{\beta}
\newcommand{\dl}{\delta}
\newcommand{\la}{\lambda}

\newcommand{\f}{\phi}

\newcommand{\p}{\pi}
\newcommand{\s}{\psi}
\newcommand{\x}{\xi}

\newcommand{\m}{\mu}
\newcommand{\n}{\nu}
\newcommand{\g}{\gamma}
\newcommand{\G}{\Gamma}
\newcommand{\e}{\epsilon}
\newcommand{\ve}{\varepsilon}
\newcommand{\th}{\theta}

\newcommand{\vt}{\vartheta}

\newcommand{\si}{\sigma}
\newcommand{\Si}{\Sigma}

\newcommand{\Y}{Y\to X}
\newcommand{\w}{\wedge}

\newcommand{\wt}{\widetilde}
\newcommand{\wh}{\widehat}
\newcommand{\ol}{\overline}

\newcommand{\dr}{\partial}

\newcommand{\ar}{\op\longrightarrow}

\newcommand{\mto}{\mapsto}

\newcommand{\ox}{\otimes}

\newcommand{\ot}{\otimes}

\let\ssection=\section
\renewcommand{\section}{\setcounter{equation}{0}\ssection}

\newcounter{eqalph}[section]
\newcounter{equationa}[section]
\newcounter{example}[section]
\newcounter{remark}[section]
\newcounter{theorem}[section]
\newcounter{proposition}[section]
\newcounter{lemma}[section]
\newcounter{corollary}[section]
\newcounter{definition}[section]

\def\theremark{\arabic{section}.\arabic{remark}}

\def\thedefinition{\arabic{section}.\arabic{definition}}

\newenvironment{theo}{\refstepcounter{definition} \bigskip\noindent{\sc
Theorem \thedefinition}.}{$\Box$\bigskip }
\newenvironment{prop}{\refstepcounter{definition} \bigskip\noindent{\sc
Proposition \thedefinition}.}{$\Box$ \bigskip }

\newenvironment{eqalph}{\stepcounter{equation}
\setcounter{equationa}{\value{equation}}
\setcounter{equation}{0}

\begin{eqnarray}}{\end{eqnarray}
\setcounter{equation}{\value{equationa}}}

\newcommand{\mar}[1]{}

\begin{document}
\hbox{}

{\parindent=0pt

{\large\bf On the geometric foundation of classical gauge gravitation theory}
\medskip

{\sc  Gennadi
Sardanashvily}
\medskip

\begin{small}
Department of Theoretical Physics, Physics Faculty, Moscow State
University, 117234 Moscow, Russia

E-mail: sard@grav.phys.msu.su

\bigskip

{\bf Abstract.}

A number of recent works in E-print arXiv have addressed the foundation
of gauge gravitation theory again. As is well known, differential
geometry of fibre bundles provides the adequate mathematical formulation
of classical field theory, including gauge theory on principal bundles.
Gauge gravitation theory is formulated on the natural bundles over a
world manifold whose structure group is reducible to
the Lorentz group. It is the metric-affine gravitation theory where
a metric (tetrad) gravitational field is a Higgs field.

\end{small}
}

\section{Introduction}

The present exposition of gauge gravitation theory follows 
Refs. \cite{book00,sard97b,sard98a}. 

By a world manifold throughout is meant a four-dimensional oriented
smooth manifold coordinated
by $(x^\la)$.

Let us first recall the notion of a gauge transformation
\cite{book,book00,mara}. In the physical literature, by a (general)
gauge transformation is meant 
a bundle automorphisms $\Phi$ of a fibre bundle $Y\to X$ over a
diffeomorphism $f$ of 
its base $X$. If $f=\id X$, the $\Phi$ is said to be a vertical gauge
transformation. If $P\to X$ is a principal bundle with a
structure Lie group $G$, a gauge transformation $\Phi$ of $P$ is an
equivariant automorphism of $P$, i.e., $\Phi\circ R_g=R_g\circ\Phi$,
$g\in G$, where $R_g$ denotes the canonical right action of $G$ on $P$
on the right. A diffeomorphism $f$ of $X$ need not give rise to an
automorphism a fibre bundle $Y\to X$, unless $Y$ belongs
to the category of natural bundles over $X$. A natural
bundle $T\to X$, by definition, admits the  
canonical lift of any diffeomorphism $f$ of its base $X$ to a bundle
automorphism $\wt f$ of $T$ over $f$ \cite{kol}. This lift $\wt f$ is called a
general covariant transformation (or a holonomic automorphism).
Natural bundles are exemplified by tensor bundles. They are associated
with the principal bundle $LX\to X$
of oriented linear frames in the tangent spaces to $X$.

The familiar gauge theory of internal symmetries deals with only
vertical gauge transformations, while generic gauge transformations in
gravitation theory are
general covariant transformations. Therefore, gauge gravitation theory
cannot straightforwardly follow the scheme of gauge theory of internal
symmetries. 
For instance, connections on a principal bundle $P$ with the structure
group of internal symmetries $G$ are usually called the gauge
potentials of the group $G$. In gauge gravitation theory, this
terminology leads to confusion. Gauge gravitation theory deals with
connections on the principle frame bundle $LX\to X$
over a world manifold $X$. Its 
structure group is $GL_4=GL^+(4,\Bbb R)$. However, gravitation theory is not
the gauge theory of this group because gravitation
Lagrangians need not be invariant under
non-holonomic (e.g., vertical) $GL_4$-gauge transformations.

In Utiyama's seminal paper \cite{uti}, only vertical Lorentz gauge
transformations were considered. Therefore tetrad fields were
introduced by hand. 
To overcome this inconsistency, T.Kibble, D.Sciama and others
treated tetrad fields as gauge potentials of the translation group 
\cite{bas,heh76,kaw,kibbl,lord,norr,sci,tseit}. In the first variant
of this approach \cite{kibbl,sci}, the canonical lift of vector
fields on a world manifold $X$ onto a tensor bundle (i.e., generators of
one-dimensional 
groups of general covariant transformations) was falsely interpreted as
infinitesimal gauge translations. Later, the horizontal lift of vector
fields by a linear world connection was treated in the same manner
\cite{heh76}. Afterwards, the conventional gauge theory on affine fibre
bundles was called into play. However, the translation part of an
affine connection on a tangent bundle is a basic soldering form
\mar{mos035}\beq
\si=\si^\al_\la(x) dx^\la\ot\dr_\al \label{mos035}
\eeq
(see Section 2) and a tetrad field (see section 3) are proved to be different
mathematical objects \cite{iva,sard83,sardz92}. 
Nevertheless,
one continues to utilize $\si^\al_\la dx^\la$ as a
non-holonomic coframe in the metric-affine gauge theory with non-holonomic
$GL(4,\Bbb R)$ gauge transformations \cite{heh,lop}. 
On another side, 
translation gauge potentials
describe an elastic distortion in the gauge
theory of dislocations in continuous media \cite{kad,mal} and in the
analogous gauge model of the fifth force \cite{sard87,sard90,sardz92}.

For the first time, the conception of a graviton as a Goldstone
particle corresponding to breaking of Lorentz symmetries in a curved
space-time was declared at the beginning of the 1960s by Heisenberg
and D.Ivanenko. This idea was revived in the framework of constructing
non-linear (induced) representations of the group $GL(4,\Bbb R)$,
containing the Lorentz group as a Cartan subgroup \cite{ish,ogi}. 
However, the geometric aspect of gravity was not considered. The fact
that, in the framework of the gauge theory on fibre bundles, a
pseudo-Riemannian metric can be treated as a Higgs field responsible for
spontaneous breaking of space-time symmetries has been
pointed out by A.Trautman \cite{tra} and by the author \cite{sard80}.
Afterwards. it has been shown that this symmetry breaking issues from
the geometric 
equivalence principle and that it is also caused by
the existence of Dirac fermion matter with exact Lorentz
symmetries \cite{iva81,iva,man90,sard91,sardz92,epr94}. 

The geometric equivalence principle postulates that, with respect to
some reference frames, Lorentz invariants can be defined everywhere on
a world manifold \cite{iva}.
This principle has the adequate fibre bundle formulation. 
It requires
that the structure group $GL_4$ of natural bundles over
a world manifold $X$ is reducible
to the Lorentz group $SO(1,3)$. Let $LX\to X$ be the above mentioned
principle frame bundle. By virtue
of the well-known theorem \cite{kob} (see Section 3), 
its structure group $GL_4$ is
reducible to $SO(1,3)$ if and only if there exists a global section of the
quotient bundle 
\mar{b3203}\beq
\Si_{\rm PR}= LX/SO(1,3),\label{b3203}
\eeq
called the metric bundle. Its global sections are
pseudo-Riemannian metrics on $X$. Moreover, 
there is one-to-one correspondence between these global sections $g$ 
and the principal subbundles $L^gX$ 
of the frame bundle $LX$ with the structure group $SO(1,3)$. These
subbundles are called reduced Lorentz structures
\cite{book,gor,kob72,book00,zul}. 
The metric bundle $\Si_{\rm PR}$ (\ref{b3203})
is usually identified with an open subbundle of the tensor bundle
$\op\vee^2 TX$, and is coordinated by $(x^\la,\si^{\m\nu})$ with respect
to the holonomic frames $\{\dr_\la\}$ in $TX$.

In General Relativity, a pseudo-Riemannian metric describes a
gravitational field. Then, following the general scheme of gauge theory
with broken symmetries \cite{book,keyl,book00,nik,sard92,tra} (see
Section 3), one
can treat a metric gravitational field as a Higgs field, associated to a
reduced Lorentz structure. Principal connections on the principal frame
bundle $LX\to X$ and associated connections on the natural bundles over
$X$ (we agree to call them world connections on $X$) play the role of
gauge fields. Thus, we come to metric-affine 
gravitation theory \cite{heh,mcrea,miel,obu97}. Its
configuration space is the bundle product
$\Si_{\rm PR}\op\times_XC_K$, where 
\mar{015}\beq
C_K=J^1LX/GL_4 \label{015}
\eeq
is the fibre bundle, coordinated by $(x^\la, k_\la{}^\nu{}_\al)$, whose
sections are world connections 
\mar{B}\beq
K= dx^\la\otimes (\dr_\la +K_\la{}^\m{}_\n \dot x^\n
\dot\dr_\m) \label{B}
\eeq
on $X$ (see Section 2). Given a pseudo-Riemannian metric, any world
connection is expressed into the Christoffel symbols, torsion and
non-metricity tensors.

Generic gauge transformations of metric-affine gravitation theory 
are general covariant transformations of natural. All gravitation 
Lagrangians, by construction, are invariant under these
transformations, but need not be invariant under non-holonomic (e.g.,
vertical) $GL_4$-gauge transformations. At the same time, if a
metric-affine Lagrangian  
factorizes through the curvature 
$R_{\la\m}{}^\al{}_\bt$ of a world connection, it
is invariant under the (non-holonomic) dilatation gauge transformations
\mar{gg0}\beq
k_\m{}^\al{}_\bt\mapsto k_\m{}^\al{}_\bt + V_\m\dl^\al_\bt, \label{gg0}
\eeq
called the projective freedom. The problem is that this invariance
implies the corresponding projective freedom of a matter Lagrangian which
imposes rigorous constraints on matter sources. To avoid the projective
freedom, one usually add non-curvature terms to metric-affine
Lagrangians expressed into the irreducible parts of 
torsion and non-metricity tensors \cite{obu97}. In this case, the Proca field
\cite{der} and the hypermomentum fluid \cite{bab,obu97} can play the
role of hypothetical matter sources of non-metricity.

To dispose of the projective freedom problem, one can consider
only Lorentz connections. A Lorentz connection is a principal connection
on the frame bundle $LX$ 
which is reducible to a principal
connection on some Lorentz subbundle $L^gX$ of $LX$. This takes place iff
the covariant derivative of the metric $g$ with respect to this
connection vanishes. Conversely, any principal connection on a reduced
Lorentz subbundle is extended to a principal connection on the frame
bundle $LX$, i.e., it defines a world connection on $X$. The holonomy group of
a Lorentz connection is $SO(1,3)$. Metric-affine
gravitation theory restricted to Lorentz connections is gravitation
theory with torsion \cite{shap}. The Levi--Civita connection of a
pseudo-Riemannian 
metric $g$ is a torsion-free Lorentz connection on the reduced Lorentz
bundle $L^gX$. Considering only Levi--Civita connections, one comes to General
Relativity. 

Given a reduced Lorentz subbundle $L^gX$ and Lorentz connections on it,
one can consider an associated vector bundle with the structure Lorentz
group whose sections are treated as matter fields.
The only realistic example of these matter fields is a Dirac spinor field.
Its symmetry group $L_\rs=SL(2,\Bbb C)$ (see Section 5) is the
two-fold universal covering of the proper (connected) Lorentz group
L$=SO^0(1,3)$. Therefore, the existence of Dirac spinor fields
on a world manifold implies the contraction of the structure group $GL_4$
of the principal frame bundle $LX$ to the proper Lorentz group L. In
accordance with the above mentioned theorem, this contraction takes
place iff there exists a global section $h$ of the quotient bundle
\mar{5.15}\beq
\Si=LX/\rL,  \label{5.15}
\eeq
called the tetrad bundle. 
It is the two-fold covering of
the metric bundle $\Si_{\rm PR}$ (\ref{b3203}).
Its global sections are called tetrad
fields. A tetrad field $h$ is represented by local sections
$h^a_\la(x)$ of the corresponding L-principal subbundle $L^hX$ of the
frame bundle $LX$ (see Section 4). They are called tetrad functions. 
Obviously, every tetrad field $h:X\to\Si$ defines a unique
pseuso-Riemannian metric $g:X\to\Si_{\rm PR}$ such that the familiar relation
\mar{mos175}\beq
g_{\m\nu}=h_\m^ah_\nu^b\eta^{ab},
\label{mos175}
\eeq
where $\eta$ is the Minkowski metric, holds. Therefore, one can think
of $h$ as being a tetrad gravitational field. 

It should be emphasized that tetrad functions of a tetrad field $h$ are
defined up to vertical Lorentz gauge 
transformations of the principal bundle $L^hX$. Therefore, tetrad
gravitation theory possesses additional vertical Lorentz gauge
transformations. However, these transformations are not similar to
vertical gauge transformations in gauge theory of internal symmetries
because they do not transform a tetrad field, but change only its
representation by local tetrad functions.

To include into consideration Dirac spinor fields, one should consider 
spinor bundles over a world manifolds. 
A Dirac spin structure
on a world manifold $X$ is defined as a pair
$(P^h, z_\rs)$ of an $L_\rs$-principal bundle $P^h\to X$ and a 
bundle morphism
\mar{b3246}\beq
z_\rs: P^h \to LX \label{b3246}
\eeq
from $P^h$ to the frame bundle $LX$ \cite{avis,benn,law}. 
Every morphism
(\ref{b3246}) factorizes through a morphism 
\mar{mos250}\beq
z_h: P^h \to L^hX \label{mos250}
\eeq
of $P^h$ to some
L-principal subbundle $L^hX$ of the frame bundle $LX$.
It follows that the necessary condition for
existence of a Dirac spin structure on a world manifold $X$ is that $X$
admits a
Lorentz structure and, consequently, a gravitational field exists. 

Dirac spinor fields in the presence of a tetrad field $h$ are
described by sections of the $P^h$-associated spinor bundle $S^h\to X$.
In order to construct the Dirac operator on
this bundle, one should define: (i) the representation
\mar{gg1}\beq
dx^\la\mapsto \g_h(dx^\la)=h^\la_a(x)\g^a \label{gg1}
\eeq
of coframes $\{dx^\la\}$ by the Dirac $\g$-matrices on elements of
$S^h$ and (ii) the covariant derivative of sections of $S^h$ with
respect to a world connection on $X$ \cite{book,book00,sard97b,sard98a}.
The result is the following (see Section 5).

(i) One can show that, for different tetrad fields $h$ and $h'$, the
representations $\g_h$ and $\g_{h'}$ (\ref{gg1}) are not equivalent. This fact
exhibits the physical nature of gravity as a Higgs field
\cite{book,book00,sardz92,sard98a}. 

(ii) Given a tetrad field $h$, there is one-to-one correspondence
between Lorentz connections on the L-principal bundle $L^hX$ and the
principal connections, called spin connections, on the $h$-associated
principal spinor bundle. For instance, every Levi--Civita connection
yields the Fock--Ivanenko spin connection. Furthermore, one can show
that, though a world 
connection need not be a Lorentz connection, any world connection
yields a Lorentz connection on each L-principal subbundle $L^hX$ and,
consequently, a spin connection 
\cite{aring,book,book00,pon,sard97b}. 
It follows that gauge gravitation theory
in the presence of Dirac spinor fields comes to the metric-affine
theory, too.

This theory admits vertical spinor gauge transformations of 
Dirac spinor fields and vertical Lorentz gauge transformations of 
tetrad functions which lead to the equivalent representations (\ref{gg1}).
The problem lies in extension of general covariant transformations to
spinor fields.

The group $GL_4$ has the universal two-fold covering group $\wt{GL}_4$ such
that the diagram
\mar{b3243}\beq
\begin{array}{ccc}
 \wt{GL}_4 & \longrightarrow &  GL_4 \\
 \put(0,-10){\vector(0,1){20}} & 
& \put(0,-10){\vector(0,1){20}}  \\
L_\rs & \ar^{z_L} &  L
\end{array} \label{b3243}
\eeq
commutes. Let us consider 
the corresponding two-fold covering bundle $\wt{LX}\to X$ of the
frame bundle $LX$ \cite{dabr,law,perc,swit}. 
The group $\wt{GL}_4$ admits the spinor representation, but it is  
infinite-dimensional. As a consequence, the $\wt{LX}$-associated spinor
bundle over $X$ describes infinite-dimensional "world" spinor fields, but not
the Dirac ones \cite{heh}.
At the same time, since the fibre bundle 
\mar{mos265}\beq
\wt{LX}\to\Si \label{mos265}
\eeq
is an $L_\rs$-principal bundle over the 
tetrad bundle $\Si=\wt{LX}/L_\rs$, the commutative diagram 
\mar{b3250}\beq
\begin{array}{rcl}
 \wt{LX}  & \op\longrightarrow^{\wt z} &  LX \\
  & \searrow  \swarrow & \\ 
 & {\Si} &  
\end{array} \label{b3250}
\eeq
provides a Dirac spin structure on the tetrad bundle $\Si$.
This spin structure is unique and possesses
the following important property \cite{giach99,book00,sard98a}.

Let us consider the spinor bundle
$S\to\Si$, associated with the $L_\rs$-principal bundle (\ref{mos265}).
We have the composite bundle 
\mar{y2}\beq
S\to \Si \to X. \label{y2}
\eeq
Given a tetrad field $h$, the
restriction $h^*S$ of the spinor bundle $S\to \Si$ to $h(X)\subset
\Si$ is isomorphic to 
the $h$-associated spinor bundle $S^h\to X$ (see Section 3), i.e., its
sections are Dirac spinor fields in the presence of a tetrad
gravitational field $h$. Conversely, every Dirac spin bundle $S^h$ over
a world manifold can be obtained
in this way. The key point is that $S\to X$ is not a spinor bundle, and 
admits general covariant transformations. 
The corresponding canonical lift onto $S$ of a vector field on $X$ can be
constructed \cite{book,book00,sard98a}. The goal
is the energy-momentum conservation law in metric-affine gravitation theory 
in the presence of fermion fields.
One can show that the corresponding energy-momentum current reduces to the
generalized Komar superpotential \cite{book,book00,sard97b}. 

The work is organized as follows. Section 2 is devoted to geometry
of linear and affine frame bundles and world connections. 
Section 3 is concerned with the reduced structures and geometry of
spontaneous symmetry breaking. Section
4 addresses the reduced Lorentz structures. 
Section 5 is devoted to spin structures on a world
manifold. For the convenience of the reader, the relevant material on
geometry of fibre bundles, jet manifolds and connections 
is summarized in Sections 6 -- 7 (see \cite{book,book00} for
the detailed exposition).

\section{Frame bundles and world connections}

Let $X$ be a world manifold. Its coordinate atlas $\Psi_X$
and the corresponding holonomic bundle atlas  
$\Psi$ (\ref{mos150}) of the tangent bundle $TX$ hold fixed.

Let $\pi_{LX}:LX\to X$ be the $GL_4$-principal bundle
of oriented linear
frames in the tangent spaces to a world manifold $X$. 
Its sections are called frame fields. 
Given holonomic frames
$\{\dr_\m\}$ in the tangent bundle $TX$, every element
$\{H_a\}$ of the frame bundle
$LX$ takes the form $H_a=H^\m_a\dr_\m$,
where $H^\m_a$ is a matrix element of the natural representation of the
group $GL_4$ in $\Bbb R^4$. These matrix elements
constitute the bundle coordinates 
\be
(x^\la, H^\m_a), \qquad H'^\m_a=\frac{\dr x'^\m}{\dr x^\la}H^\la_a, 
\ee
on $LX$. In these coordinates, the canonical action  
(\ref{1}) of the structure group $GL_4$ on $LX$ reads 
\be
R_g: H^\m_a\mapsto H^\m_bg^b{}_a, \qquad g\in GL_4.
\ee
The frame bundle $LX$ is equipped with the canonical $\Bbb
R^4$-valued 1-form 
\mar{b3133'}\beq
\th_{LX} = H^a_\m dx^\m\ot t_a, \label{b3133'}
\eeq
where $\{t_a\}$ is a fixed basis for $\Bbb R^4$ and $H^a_\m$ is the inverse
matrix of $H^\m_a$.

The frame bundle $LX\to X$ belongs to the category of natural bundles.
Every diffeomorphism $f$ of $X$ gives rise to the
automorphism
\mar{025}\beq
\wt f: (x^\la, H^\la_a)\mapsto (f^\la(x),\dr_\m f^\la H^\m_a) \label{025}
\eeq
of $LX$.
The lift (\ref{025}) implies the canonical lift $\wt\tau$ of every vector
field $\tau$ on
$X$ onto the principal bundle $LX$. It is defined by the relation
\be
\bL_{\wt\tau}\th_{LX}=0,
\ee
where $\wt\tau$ (\ref{l27}) is the canonical lift of $\tau$ onto the
tangent bundle $TX$.

Every
world connection $K$ (\ref{B}) on a world manifold $X$ is associated
with a principal connection on $LX$.
Consequently, there is
one-to-one correspondence 
between the world connections and the sections of 
the quotient fibre bundle $C_K$ (\ref{015}), 
called the bundle of world connections. 
With respect to the holonomic frames in $TX$, the fibre bundle $C_K$
is provided with the bundle coordinates $(x^\la, k_\la{}^\nu{}_\al)$ such
that, for any 
section
$K$ of $C_K\to X$,
\be
k_\la{}^\nu{}_\al\circ K=K_\la{}^\nu{}_\al
\ee
are the coefficients of the world connection $K$ (\ref{B}). 
Though the bundle of world connections $C_K$ (\ref{015}) is not an
$LX$-associated bundle, it is a natural bundle and admits the
canonical lift
\mar{b3150}\beq
\wt\tau = \tau^\m\dr_\m +[\dr_\nu\tau^\al k_\m{}^\nu{}_\bt -
\dr_\bt\tau^\nu k_\m{}^\al{}_\nu - \dr_\m\tau^\nu
k_\nu{}^\al{}_\bt + \dr_{\m\bt}\tau^\al]\frac{\dr}{\dr k_\m{}^\al{}_\bt}
\label{b3150}
\eeq
of any vector field $\tau$ on $X$ \cite{book,book00}.

The torsion of a world connection is the
vertical-valued 2-form $T$ (\ref{191}) on $TX$. 
Due to the canonical vertical splitting (\ref{mos163}), the
torsion (\ref{191}) is represented by the tangent-valued 2-form (\ref{1190})
on $X$ and the soldering form 
\mar{mos160}\beq
T=T_\m{}^\n{}_\la \dot x^\la dx^\m\ot\dot \dr_\nu. \label{mos160}
\eeq

Every world connection $K$ yields the horizontal lift 
\mar{b3180}\beq
K\tau =\tau^\la(\dr_\la +K_\la{}^\bt{}_\al\dot x^\al\dot\dr_\bt)
\label{b3180}
\eeq
of a vector field
$\tau$ on $X$ onto $TX$. 
One can think of this lift as being the generator of a
local 1-parameter group of non-holonomic automorphisms of
this bundle. Note that, in the pioneer gauge gravitation models, the
canonical lift (\ref{l27}) and the horizontal lift (\ref{b3180}) were treated
as generators of the gauge group of 
translations \cite{heh76,iva} (see \cite{maar} for some other lifts
onto $TX$ of vector 
fields on $X$).

Being a vector bundle, the tangent bundle $TX$ over a world manifold has a
natural structure of an affine bundle. Therefore, one can consider affine
connections on $TX$, called affine world connections. 
Let us study them as principal connections.

Let $Y\to X$ be an affine bundle with a $k$-dimensional typical fibre
$V$. It is associated with a principal bundle $AY$ of affine frames in $Y$
whose structure group is the general affine group $GA(k,\Bbb R)$. Then any
affine connection on $Y\to X$ can be seen as an associated with a principal
connection on $AY\to X$. These connections are represented by global
sections of 
the affine bundle $J^1P/GA(k,\Bbb R)\to X$.

Every affine connection $\G$ (\ref{184}) on
$Y\to X$ defines a linear connection $\ol G$ (\ref{mos032}) on the underlying
vector bundle
$\ol Y\to X$. This connection $\ol\G$ is associated with a linear principal
connection  on the principal bundle $L\ol Y$ of linear frames in $\ol Y$
whose structure group is the general linear group $GL(k,\Bbb R)$. We have the
exact sequence of groups
\mar{mos031}\beq
0\to T_k\to GA(k,\Bbb R) \to GL(k,\Bbb R)\to {\bf 1}, \label{mos031}
\eeq
where $T_k$ is the group of translations in $\Bbb R^k$. It is readily
observed that there is the corresponding principal bundle morphism $AY\to
L\ol Y$ over $X$, and the principal connection $\ol G$ on $L\ol Y$ is the image
of the principal connection $\G$ on $AY\to X$ under this morphism in
accordance with Theorem \ref{mos253}. 

The exact sequence (\ref{mos031}) admits a splitting $GL(k,\Bbb
R)\hookrightarrow GA(k,\Bbb R)$, but 
one usually loses sight of the fact that this splitting is not canonical
(see, e.g., \cite{kob}). It depends on the morphism 
\be
V\ni v\mapsto v - v_0\in \ol V,
\ee 
i.e., on the choice of an origin $v_0$ of the affine space $V$. Given $v_0$,
the image of the corresponding monomorphism $GL(k,\Bbb
R)\hookrightarrow GA(k,\Bbb R)$ is the stabilizer $G(v_0)\subset GA(k,\Bbb R)$
of $v_0$.  Different subgroups $G(v_0)$ and $G(v'_0)$ are related to each
other as follows: 
\be
G(v'_0)=T(v'_0-v_0)G(v_0)T^{-1}(v'_0-v_0),
\ee
where $T(v'_0-v_0)$ is the translation along the vector $(v'_0-v_0)\in\ol V$.

Note that
the well-known morphism of a $k$-dimensional affine space $V$ onto
a hypersurface
$\ol y^{k+1}=1$ in $\Bbb R^{k+1}$ and the corresponding representation of
elements of
$GA(k,\Bbb R)$ by particular $(k+1)\times (k+1)$-matrices also fail to be
canonical. They depend on a point $v_0\in V$ sent to
vector $(0,\ldots,0,1)\in \Bbb R^{k+1}$.

If $Y\to X$ is a vector bundle, it is provided with the
canonical structure of an affine bundle whose origin is the canonical zero
section $\wh 0$. In this case, we have the canonical splitting of the exact
sequence (\ref{mos031}) such that $GL(k,\Bbb R)$ is a subgroup of $GA(k,\Bbb
R)$ and $GA(k,\Bbb R)$ is the semidirect product of $GL(k,\Bbb R)$ and the
group $T_k$ of translations in $\Bbb R^k$. Given a $GA(k,\Bbb
R)$-principal bundle $AY\to X$, its affine structure group $GA(k,\Bbb R)$ is
always reducible to the linear subgroup since the quotient $GA(k,\Bbb
R)/GL(k,\Bbb R)$ is a vector space $\Bbb R^k$ provided with the natural
affine structure. The corresponding quotient bundle is isomorphic to the
vector bundle $Y\to X$. There is the canonical injection of the linear frame
bundle $LY\hookrightarrow AY$ onto the reduced $GL(k,\Bbb R)$-principal
subbundle of $AY$ which corresponds to the zero section $\wh 0$ of $Y\to X$.
In this case, every principal connection on the linear frame bundle $LY$
gives rise to a principal connection on the affine frame bundle in
accordance with Theorem \ref{mos176}. It follows that any
affine connection $\G$ on a vector bundle $Y\to X$ defines a linear connection
$\ol\G$ on
$Y\to X$ and that every linear connection on $Y\to X$ can be seen as an affine
one. Hence, any affine connection $\G$ on the vector bundle
$Y\to X$ falls into the sum of the associated linear connection $\ol\G$ 
and a basic soldering form $\si$ on $Y\to X$.  Due to the vertical splitting
(\ref{1.10}), this soldering form is represented by a global section of the
tensor product $T^*X\ot Y$.

Let now $Y\to X$ be the tangent bundle $TX\to X$ considered as an affine
bundle. Then the relationship between affine and linear world connections on
$TX$ is the repetition of that we have said above. In particular, any
affine world connection  
\mar{mos033}\beq
K= dx^\la\ot(K_\la{}^\al{}_\m(x) \dot x^\m +\si^\al_\la(x))\dr_\al
\label{mos033}
\eeq
 on 
$TX\to X$ is represented by the sum of the associated linear world connection
\mar{mos034}\beq
\ol K= K_\la{}^\al{}_\m(x) \dot x^\m dx^\la\ot\dr_\al \label{mos034}
\eeq
on $TX$ and a basic soldering form $\si$ (\ref{mos035})
 on $TX$. For instance, if
$\si=\th_X$, we have the Cartan connection (\ref{b1.97}). 

As was mentioned above, the tensor field $\si$
(\ref{mos035}) was falsely identified with a tetrad field in some gauge
gravitation models.

\section{Geometry of spontaneous symmetry breaking}

Spontaneous symmetry breaking is a
quantum phenomenon. In classical field theory, spontaneous symmetry breaking
is modelled by classical Higgs fields. In gauge theory on a principal bundle
$P\to X$, a symmetry breaking is said to 
occur when the structure group
$G$ of $P$ is reducible to a closed subgroup $H\subset G$ of 
exact symmetries \cite{book,iva,keyl,book00,nik,sard92,tra}. 
This structure group reduction takes place iff a global
section $h$ of the quotient bundle $P/H\to X$ exists (see Theorem
\ref{redsub} below). In gauge theory, such a global section $h$ is
treated as a Higgs field. From the mathematical viewpoint, one speaks on
the Klein--Chern geometry and the reduced
$G$-structure \cite{zul}.

Let $\pi_{PX}:P\to X$
be a $G$-principal bundle (see Section 7) and  $H$ a closed 
subgroup of $G$. It is also a Lie group. We assume that dim$H>0$. 
We have the composite fibre bundle
\mar{b3223a}\beq
P\to P/H\to X \label{b3223a}
\eeq
(see Section 7), where
\mar{b3194}\beq
 P_\Si=P\ar^{\pi_{P\Si}} P/H \label{b3194}
\eeq
is a principal bundle with the structure group $H$ and
\mar{b3193}\beq
\Si=P/H\ar^{\pi_{\Si X}} X \label{b3193}
\eeq
is a $P$-associated fibre bundle with the typical fibre $G/H$ on which the
structure group $G$ acts naturally on the left. Note that the canonical
surjection $G\to G/H$ is an $H$-principal bundle.

One says that the structure group $G$ of a principal bundle $P$ is
reducible to a Lie subgroup
$H$ if there exists a $H$-principal
subbundle
$P^h$ of $P$ with the structure group $H$. This subbundle is called a 
reduced $G^\downarrow H$-structure \cite{book,gor,kob72,book00,zul}. 

Two reduced $G^\downarrow H$-structures $P^h$ and $P^{h'}$ of a $G$-principal
bundle $P$ are said to be isomorphic if there is an automorphism $\Phi$ of
$P$ which provides an isomorphism of $P^h$ and $P^{h'}$. If
$\Phi$ is a vertical automorphism of $P$, reduced structures $P^h$ and
$P^{h'}$ are called equivalent. 
Note that, in \cite{gor,kob72} (see also \cite{cord}), only reduced
structures of
the frame bundle $LX$ are considered, and a class of
isomorphisms of such reduced structures is restricted to holonomic
automorphisms of $LX$.

There are the following two
theorems \cite{kob}. 
 
\begin{theo} \label{mos96} \mar{mos96}
A structure group
$G$ of a principal bundle $P$ is reducible to its closed subgroup $H$ iff
$P$ has an atlas $\Psi_P$ with $H$-valued transition functions. 
\end{theo}

Given a reduced
subbundle $P^h$ of $P$, such an atlas $\Psi_P$ is defined by a family of local
sections $\{z_\al\}$ which take their values into $P^h$ (see Section 7).

\begin{theo}\label{redsub} \mar{redsub}
There is one-to-one correspondence $P^h=\pi_{P\Si}^{-1}(h(X))$
between the
reduced $H$-principal subbundles
$P^h$ of $P$ and the global sections $h$ of the quotient fibre bundle
$P/H\to X$ (\ref{b3193}).
\end{theo}

Given such a section $h$, let us consider 
the restriction $h^*P_\Si$ (\ref{S10}) of the
$H$-principal bundle
$P_\Si$ (\ref{b3194}) to $h(X)\subset \Si$. This is a $H$-principal bundle
over $X$ \cite{kob}, which is equivalent to the reduced
subbundle
$P^h$ of $P$.   

In general, there are topological obstructions to the reduction of a
structure group of a principal bundle to its subgroup. In accordance with
Theorem
\ref{mos9}, the structure group $G$ of a principal bundle $P$ is always
reducible 
to its closed subgroup $H$ if the quotient $G/H$ is homeomorphic to an
Euclidean space $\Bbb R^k$.

\begin{theo} \label{comp} \cite{ste}. \mar{comp} 
A structure group $G$ of a principal bundle is always reducible
to its maximal compact subgroup $H$ since the quotient space $G/H$ is
homeomorphic to an Euclidean space. 
\end{theo}

Two $H$-principal subbundles $P^h$ and $P^{h'}$ of a
$G$-principal bundle $P$ need not be isomorphic. 

\begin{prop}\label{iso1} \cite{book00}. \mar{iso1} 
(i) Every vertical automorphism $\Phi$ of the
principal bundle
$P\to X$ sends an $H$-principal subbundle $P^h$ onto an equivalent
$H$-principal subbundle
$P^{h'}$. 
(ii) Conversely, let two reduced subbundles $P^h$ and $P^{h'}$ of a principal
fibre bundle $P$ be isomorphic to each other, and
$\Phi:P^h\to P^{h'}$ be an isomorphism. Then $\Phi$ is extended to 
a vertical automorphism of $P$. 
\end{prop}

\begin{prop}\label{iso0} \cite{ste}. \mar{iso0} If the quotient $G/H$
is homeomorphic to an 
Euclidean space $\Bbb R^k$, all
$H$-principal subbundles of a $G$-principal bundle $P$ are equivalent to each
other. 
\end{prop}

Given a reduced subbundle $P^h$ of a principal bundle $P$, let
\mar{b3.3000}\beq
Y^h=(P^h\times V)/H \label{b3.3000}
\eeq
be the associated fibre bundle with a typical fibre $V$ (see Section 7).
Let $P^{h'}$ be another reduced subbundle of $P$ which is isomorphic to
$P^h$, and
\be
Y^{h'}=(P^{h'}\times V)/H
\ee
The fibre bundles $Y^h$ and $Y^{h'}$ are isomorphic, but not canonically
isomorphic in general.

\begin{prop}\label{iso2} \cite{book00}. \mar{iso2}  Let $P^h$ be an
$H$-principal subbundle of a 
$G$-principal bundle $P$. Let $Y^h$ be the $P^h$-associated bundle
(\ref{b3.3000}) with a typical fibre $V$. If $V$
carries a representation of the whole group $G$, the fibre bundle $Y^h$ is
canonically isomorphic to the $P$-associated
fibre bundle
\be
Y=(P\times V)/G.
\ee
\end{prop}

In accordance with Theorem \ref{redsub}, 
the set of reduced $H$-principal subbundles
$P^h$ of $P$ is in
bijective correspondence with
the set of Higgs fields $h$. Given such a subbundle $P^h$, let 
$Y^h$ (\ref{b3.3000}) 
be the associated vector bundle with a typical fibre $V$ which admits a
representation of the group $H$ of exact symmetries, but not of the whole
symmetry group $G$. Its sections $s_h$ describe matter fields in the
presence of the 
Higgs fields $h$ and some principal connection $A_h$ on $P^h$.
In general, the fibre bundle
$Y^h$ (\ref{b3.3000}) is not associated with other
$H$-principal subbundles $P^{h'}$ of $P$. It follows that, in this case,
$V$-valued matter fields can be represented only by pairs with 
 Higgs fields. The goal is to describe the totality of these pairs
$(s_h,h)$ for all Higgs fields $h$.

For this purpose, let us
consider the composite fibre bundle (\ref{b3223a}) and the composite
fibre bundle 
\mar{b3225}\beq
Y\ar^{\pi_{Y\Si}} \Si\ar^{\pi_{\Si X}} X \label{b3225}
\eeq
where $Y\to \Si$ is a vector bundle 
\be
Y=(P\times V)/H 
\ee 
associated with the corresponding $H$-principal bundle $P_\Si$ (\ref{b3194}). 
Given a global section $h$ of the fibre bundle $\Si\to X$ (\ref{b3193})
and the $P^h$-associated fibre bundle (\ref{b3.3000}), there is the
canonical injection 
\be
i_h: Y^h=(P^h\times V)/H \hookrightarrow Y
\ee
over $X$ whose image is the restriction
\be
h^*Y=(h^*P\times V)/H
\ee
of the fibre bundle $Y\to\Si$ to $h(X)\subset \Si$, i.e.,
\mar{b3226}\beq
i_h(Y^h)\cong \pi^{-1}_{Y\Si}(h(X)) \label{b3226}
\eeq
(see Proposition \ref{comp10}).
Then, by virtue of Proposition \ref{mos61}, every global section $s_h$
of the fibre bundle $Y^h$ corresponds to the 
global section $i_h\circ s_h$ of the composite fibre bundle (\ref{b3225}).
Conversely, every global section $s$ of the composite fibre bundle
(\ref{b3225}) which projects onto a section $h=\pi_{Y\Si}\circ s$ of the
fibre bundle $\Si\to X$ takes its values into the subbundle $i_h(Y^h)\subset
Y$ in accordance with the relation (\ref{b3226}). Hence, there is 
one-to-one correspondence between the sections of the fibre bundle $Y^h$
(\ref{b3.3000}) and the sections of the composite fibre bundle (\ref{b3225})
which cover $h$.

Thus, it is the composite fibre bundle (\ref{b3225}) whose sections
describe the above-mentioned totality of pairs $(s_h, h)$ of matter fields
and Higgs fields in gauge theory with broken symmetries
\cite{book,book00,sard92}. 

Turn now to the properties of connections compatible with a reduced
structure \cite{kob}.

\begin{theo} \label{mos176} \mar{mos176} 
Since principal connections are equivariant, every principal connection
$A_h$ on a reduced 
$H$-principal subbundle
$P^h$ of a $G$-principal bundle $P$ gives rise to a principal connection on
$P$.
\end{theo}

\begin{theo} \label{mos177} \mar{mos177}
A principal connection $A$ on a $G$-principal bundle $P$ is reducible to a
principal connection on a reduced $H$-principal subbundle $P^h$ of $P$
iff the 
corresponding global section $h$ of the $P$-associated fibre bundle
$P/H\to X$ is an integral section of the associated principal
connection $A$ on $P/H\to X$.
\end{theo}

\begin{theo} \label{mos178} \mar{mos178}
Given the composite fibre bundle (\ref{b3223a}), let $A_\Si$ be a
principal connection on the $H$-principal bundle $P\to P/H$. Then,
for any reduced $H$-principal subbundle $P^h$ of $P$, the pull-back
connection $i_h^*A_\Si$ (\ref{mos83}) is a principal connection on $P^h$.
\end{theo}

This assertion is a corollary of Theorem \ref{mos252}.

As a consequence of Theorem \ref{mos178}, there is the following
peculiarity of the dynamics of field systems with symmetry breaking.
Let the composite fibre bundle $Y$ (\ref{b3225}) be provided with coordinates
$(x^\la,
\si^m, y^i)$, where $(x^\la, \si^m)$ are bundle coordinates on the fibre
bundle $\Si\to X$.  Let 
\mar{b3228}\beq
A_\Si=dx^\la\ot(\dr_\la+ A^i_\la\dr_i)
+d\si^m\ot(\dr_m+A^i_m\dr_i) \label{b3228}
\eeq
be a principal connection on the vector bundle $Y\to \Si$.
This connection defines the splitting (\ref{46a}) of the vertical tangent
bundle $VY$ and leads to the vertical covariant differential
(\ref{7.10}) which reads
\mar{7.101}\beq
\wt D=dx^\la\ot(y^i_\la- A^i_\la -A^i_m\si^m_\la)\dr_i.\label{7.101}
\eeq
The operator (\ref{7.101}) possesses the following property (see
Section 7).
Given a global section $h$ of $\Si\to X$, its restriction 
\mar{b3260}\ben
&&\wt D_h =\wt D\circ J^1i_h: J^1Y^h \to T^*X\ot VY^h, \label{b3260}\\
&& \wt D_h =dx^\la\ot(y^i_\la- A^i_\la -A^i_m\dr_\la h^m)\dr_i, \nonumber
\een
to $Y^h$ is the familiar covariant differential relative to the 
pull-back principal connection $A_h$ (\ref{mos83}) 
on the fibre bundle $Y^h\to X$. Thus, a Lagrangian  on the
jet manifold $J^1Y$ of a composite fibre bundle usually factorizes through the
vertical differential $\wt
D_A$ \cite{book,book00}.

\section{Lorentz structures}

Let us apply the above scheme of symmetry breaking in gauge theory to
gravitation theory.

As was mentioned above, the geometric formulation of the equivalence principle 
requires that the
structure group 
$GL_4$ of the frame bundle $LX$ over a world manifold $X$ is reducible
to the Lorentz group $SO(1,3)$, while the condition of existence of fermion
fields implies that $GL_4$ is reducible to the proper Lorentz group
$\rL$.  The latter is homeomorphic to ${\bf RP}^3\times
\Bbb R^3$, where ${\bf RP}^3$ is a real 3-dimensional projective space.
Unless otherwise stated, by a
Lorentz structure we will mean 
a reduced L-principal subbundle $L^hX$ of $LX$
called the Lorentz
subbundle.

Note that there is the topological obstruction to the existence of a
reduced Lorentz structure on a world manifold $X$. 
All non-compact manifolds and compact manifolds whose Euler characteristic 
equals zero admit a reduced $SO(1,3)$-structure and, as a
consequence, a pseudo-Riemannian metric \cite{dods}. A
reduced L-structure exists if $X$ is additionally time-orientable.
In gravitational models,
certain conditions of causality should also be satisfied (see \cite{haw}). 
A compact space-time does not possess this property. At the same time,  
a non-compact world manifold $X$ has a spin
structure iff it is parallelizable (i.e., the tangent bundle $TX\to
X$ is trivial) \cite{ger,wist}. 

Let us assume that a Lorentz structure on a world
manifold exists. Then one can show that different Lorentz subbundles
$L^hX$ and $L^{h'}X$ of
the frame bundle $LX$ are equivalent as L-principal bundles
\cite{henn}. It means that there exists a vertical automorphism of the
frame bundle 
$LX$ which sends isomorphically $L^hX$ onto $L^{h'}X$ (see Proposition
\ref{iso1}). 

By virtue of Theorem \ref{redsub}, there is
one-to-one correspondence between the L-principal subbundles
$L^hX$ of
the frame bundle  $LX$  and the global
sections $h$ of the tetrad bundle $\Si$ (\ref{5.15})
This is an $LX$-associated fibre bundle with the typical 
fibre $GL_4/$L, homeomorphic to $S^3\times\Bbb R^7$. 
Its global sections are tetrad fields. Note that the typical fibre of
the metric bundle $\Si_{\rm PR}$ (\ref{b3203}) is homeomorphic to 
${\bf RP}^3 \times \Bbb R^7$.

Every tetrad field $h$ defines an associated 
Lorentz atlas
$\Psi^h=\{(U_\zeta,z_\zeta^h)\}$ of the frame bundle
$LX$ such that the corresponding local sections $z_\zeta^h$ of 
$LX$ take their values into the Lorentz subbundle $L^hX$. They are
called tetrad functions, and are given by the coordinate expression 
\mar{L6}\beq
h^\m_a =H^\m{}_a\circ z^h_\zeta. \label{L6}
\eeq
In accordance
with Theorem (\ref{mos96}) the transition functions of the Lorentz
atlases of the frame bundle $LX$ and associated bundles are L-valued.

Given a Lorentz atlas $\Psi^h$, the
pull-back $z_\zeta^{h*}\th_{LX}$
of the canonical form $\th_{LX}$ (\ref{b3133'}) by a tetrad 
function 
$z_\zeta^h$ is called a (local) tetrad form. 
It reads 
\mar{b3211}\beq
h^a\ot t_a=z_\zeta^{h*}\th_{LX}=h_\la^a dx^\la\ot t_a, \label{b3211}
\eeq
where $h_\la^a$ are elements of the inverse matrix to $h^\m_a$ (\ref{L6}).
The tetrad form (\ref{b3211}) determines the tetrad
coframes
\mar{b3211'}\beq
h^a =h^a_\m(x)dx^\m, \qquad x\in U_\zeta, \label{b3211'}
\eeq
in the cotangent bundle $T^*X$. These coframes are associated with the
Lorentz atlas $\Psi^h$.
In particular, the relation (\ref{mos175}) 
between the tetrad functions and the metric functions of the corresponding
pseudo-Riemannian metric $g:X\to\Si_{\rm PR}$
takes the form
\be
g=\eta_{ab}h^a\ot h^b.
\ee
It follows that this pseudo-Riemannian metric $g$
has the Minkowski metric functions with
respect to any Lorentz atlas $\Psi^h$. It exemplifies a Lorentz
invariant mentioned in the geometric equivalence principle.

Let $M=\Bbb R_4$ be the dual of $\Bbb R^4$ provided  
with the Minkowski metric $\eta$. Given a tetrad field $h$, let 
us consider the $L^hX$-associated fibre bundle  of
Minkowski spaces 
\mar{b3192}\beq
M^hX=(L^hX\times M)/\rL. \label{b3192}
\eeq
By virtue of Proposition \ref{iso2}, it is isomorphic to the cotangent
bundle 
\mar{b3.4000}\beq
T^*X=(LX\times\Bbb R_4)/GL_4= (L^hX\times M)/\rL=M^hX. \label{b3.4000}
\eeq
Given the isomorphism (\ref{b3.4000}), we say that the cotangent bundle
$T^*X$ is provided
with a Minkowski structure. 
Note that different Minkowski structures $M^hX$ and $M^{h'}X$ on $T^*X$
are not equivalent.

As was mentioned above, a principal connection on a Lorentz subbundle
$L^h$ of the frame bundle 
$LX$ is called the Lorentz connection. 
It reads
\mar{b3205}\beq
A_h=dx^\la\ot(\dr_\la + \frac12A_\la{}^{ab} \ve_{ab}) 
\label{b3205}
\eeq
where $\ve_{ab}=-\ve_{ba}$ are generators of the Lorentz group.
Recall that the Lorentz group L acts on $\Bbb R^4$ by the generators
\mar{b3278}\beq
\ve_{ab}{}^c{}_d= \eta_{ad}\dl^c_b- \eta_{bd}\dl^c_a. \label{b3278}
\eeq
By virtue of Theorem \ref{mos176}, every Lorentz connection
(\ref{b3205}) is extended to a principal connection on the frame bundle
$LX$ and, thereby,
it defines a world connection $K$ whose coefficients are
\mar{mos190}\beq
K_\la{}^\m{}_\nu = h^k_\nu\dr_\la h^\m_k + \eta_{ka}h^\m_b h^k_\nu
A_\la{}^{ab}. \label{mos190}
\eeq
This world connection is also called the Lorentz connection. Its
holonomy group is a subgroup of the Lorentz group L. Conversely, let $K$ be a
world connection with the holonomy group $L$. By virtue of the well-known
theorem \cite{kob},
it defines a Lorentz subbundle of the frame bundle $LX$,
and is a Lorentz connection on this subbundle. 

Though a world connection is not a Lorentz connection in general,
any world connection $K$ defines a Lorentz connection $K_h$ on each
$L$-principal subbundle $L^hX$ of the frame bundle as follows.

It is readily observed that the Lie algebra of the general linear group
$GL_4$ is the direct sum
\be
{\got g}(GL_4) = {\got g}(\rL) \oplus {\got m}
\ee
of the Lie algebra ${\got g}(\rL)$ of the Lorentz group and a subspace ${\got
m}$ such that 
$ad(l)({\got m})\subset {\got m}$ for all $l\in \rL$.
Let $\ol K$ be the connection form of a principal
connection $K$ on $LX$. Then, by virtue of the well-known theorem \cite{kob},
the pull-back on $L^hX$ of the ${\got g}(\rL)$-valued component $\ol K_L$
of
$\ol K$ is the connection form  of a principal connection
$K_h$ on the 
Lorentz subbundle $L^hX$. To obtain the
connection parameters of $K_h$, 
let us consider the local connection 1-form of the
connection $K$ with 
respect to a Lorentz atlas $\Psi^h$ of $LX$ given by the tetrad forms
$h^a$. This reads
\be
z_\zeta^{h*}\ol K=- K_\la{}^b{}_a dx^\la\ot \ve_b^a,\qquad
K_\la{}^b{}_a = -h^b_\m \dr_\la h^\m_a  + K_\la{}^\m{}_\nu h^b_\m
h^\nu_a,
\ee
where $\{\ve_b^a\}$ is the basis for the right Lie algebra of the group $GL_4$.
Then, the Lorentz part of this form is the
local connection 1-form of the connection $K_h$ on
$L^hX$. We have 
\mar{K102}\beq
z_\zeta^{h*}\ol K_L= -\frac12 A_\la{}^{ab}dx^\la\ot \ve_{ab},  \qquad
 A_\la{}^{ab} =\frac12 (\eta^{kb}h^a_\m-\eta^{ka}h^b_\m)(\dr_\la h^\m_k -
 h^\nu_k K_\la{}^\m{}_\nu). \label{K102}
\eeq
If $K$ is a Lorentz connection extended from $L^hX$, then obviously $K_h=K$.

\section{Dirac spin structures}

We describe Dirac spinors in terms of Clifford algebras
(see, e.g., \cite{budi,cra,law,obu,rodr}).

Let $M$ be the Minkowski space equipped with the Minkowski metric $\eta$,
and $\{e^a\}$ be a fixed basis for $M$. By
$\Bbb C_{1,3}$ is denoted the complex Clifford
algebra generated by elements of
$M$. This is the complexified quotient of
the tensor algebra $\ot M$ of $M$ by the two-sided ideal generated by
elements
\be
e\otimes e'+e'\otimes e-2\eta(e,e')\in \ot M,\qquad e,e'\in M.
\ee
The complex Clifford algebra $\Bbb C_{1,3}$ is isomorphic to the real
Clifford algebra $\Bbb R_{2,3}$, whose generating
space is $\Bbb R^5$ equipped with the metric 
\be
{\rm diag}(1,-1,-1,-1,\,1).
\ee
Its
subalgebra generated by elements of $M\subset \Bbb
R^5$ is the real Clifford algebra 
$\Bbb R_{1,3}$. 

A spinor space $V$ is defined as a
minimal left ideal of $\Bbb C_{1,3}$  on
which this algebra acts on the left. We have the representation
\mar{w01}\beq
\g: M\otimes V \to V, \qquad \g(e^a)=\g^a, \label{w01}\\
\eeq
of elements of the Minkowski space $M\subset{\Bbb C}_{1,3}$ by the Dirac
$\g$-matrices on $V$. Different ideals $V$ lead to
equivalent representations (\ref{w01}).

By definition, the Clifford group $G_{1,3}$
consists of the invertible elements $l_s$ of
the real Clifford algebra $\Bbb R_{1,3}$ such that the inner automorphisms
defined by these elements preserve the Minkowski space $M\subset \Bbb
R_{1,3}$, that is,
\mar{b3200}\beq
l_sel^{-1}_s = l(e), \qquad e\in M, \label{b3200}
\eeq
where $l$ is a Lorentz transformation of $M$. 
Hence, we have an
epimorphism of the Clifford group $G_{1,3}$ onto the Lorentz group $O(1,3)$.
Since the action (\ref{b3200}) of the Clifford group on the Minkowski space 
$M$ is not effective, one usually consider its
pin and spin subgroups.  
The 
subgroup $Pin(1,3)$ of $G_{1,3}$ is generated by elements $e\in
M$ such that $\eta(e,e)=\pm 1$. The even part of 
Pin$(1,3)$ is the spin group $Spin(1,3)$.
Its component of the unity is the above mentioned group 
two-fold universal covering group
\mar{b3204}\beq
z_L:L_\rs\to \rL=L_\rs/\Bbb Z_2, \qquad \Bbb Z_2=\{1,-1\}, \label{b3204}
\eeq
of the proper Lorentz group L.

The Clifford group $G_{1,3}$ acts on the spinor space $V$ by left
multiplications
\be
G_{1,3}\ni l_s:v\mapsto l_sv, \qquad v\in V. 
\ee
This action preserves the
representation (\ref{w01}), i.e.,
\be
\g (lM\otimes l_sV) = l_s\g (M\otimes V). 
\ee
The spin group $L_{\rs}$ acts on the spinor space $V$
by means of the generators
\mar{b3213}\beq
L_{ab}=\frac{1}{4}[\g_a,\g_b]. \label{b3213}
\eeq

Let $P^h\to X$ be a principal bundle with the structure group $L_\rs$
provided with the bundle morphisms $z_\rs$ (\ref{b3246}) and $z_h$
(\ref{mos250}). Dirac spinor fields in the presence of a tetrad field $h$ are
described by sections of the $P^h$-associated spinor bundle
\mar{y1}\beq
S^h=(P^h\times V)/L_\rs\to X, \label{y1}
\eeq
whose typical fibre $V$ carriers the spinor representation
(\ref{b3213}) of the spin group $L_\rs$. To define the Dirac operator
on this fields, the
spinor bundle $S^h$ (\ref{y1}) must be represented as a subbundle of
the bundle of Clifford algebras, i.e., as a spinor structure on the
cotangent bundle $T^*X$. 

Every fibre bundle of 
Minkowski spaces $M^hX$ (\ref{b3192}) over a world manifold $X$ 
is extended to the fibre bundle of Clifford algebras $C^hX$ 
with the fibres 
generated by the fibres of $M^hX$ \cite{benn}. 
This fibre bundle $C^hX$ has the structure 
group Aut$(\Bbb C_{1,3})$ of inner automorphisms of the Clifford 
algebra $\Bbb C_{1,3}$.
In general, $C^hX$ does not 
contain a spinor subbundle because a 
spinor subspace $V$ 
is not stable under inner automorphisms of $\Bbb C_{1,3}$. 
As was shown \cite{benn}, a spinor subbundle of $C^hX$ exists if
the transition functions of $C^hX$ 
can be lifted from Aut$(\Bbb C_{1,3})$ to the Clifford group $G_{1,3}$. 
This agrees with 
the standard condition of existence of a spin structure on a
world manifold $X$.
Such a spinor subbundle is the bundle $S^h$ (\ref{y1}) 
associated with the universal two-fold
covering (\ref{mos250}) of the Lorentz bundle $L^hX$. We will call
$P^h$ (and $S^h$) the $h$-associated Dirac spin structure 
on a world manifold. 

Note that all spin structures on a manifold $X$ which
are related to the two-fold universal covering groups possess the following
two properties \cite{greub}.
Let $P\to X$ be a principal bundle whose structure group $G$ has the
fundamental group $\pi_1(G)=\Bbb Z_2$. Let $\wt G$ be the universal covering
group of $G$.
\begin{itemize}
\item The topological obstruction to the existence of a 
$\wt G$-principal bundle $\wt P\to X$ covering the bundle $P\to X$
is given by the \v Cech
cohomology group
$H^2(X;\Bbb Z_2)$ of $X$ with coefficients in $\Bbb Z_2$. Roughly speaking,
the principal bundle $P$ defines an element of $H^2(X;\Bbb Z_2)$ which must
be zero so that $P\to X$ can give rise to $\wt P\to X$.
\item Non-equivalent lifts of $P\to X$ to  $\wt G$-principal bundles
are classified by elements of the \v Cech cohomology group $H^1(X;\Bbb Z_2)$.
\end{itemize}
In particular, the well-known topological obstruction to the existence of a
Dirac spin structure is the  second Stiefel--Whitney  
class
$w_2(X)\in H^2(X;\Bbb Z_2)$  of $X$ \cite{law}.
In the case of 4-dimensional non-compact manifolds,
all Riemannian and pseudo-Riemannian spin structures are equivalent
\cite{avis,ger}.

There exists the bundle morphism
\mar{L4}\beq
\g_h: T^*X\ot S^h=(P^h\times (M\ot V))/L_\rs\to (P^h\times
\g(M\ot V))/L_\rs=S^h, \label{L4}
\eeq
where by $\g$ is meant the left action (\ref{w01}) of $M\subset \Bbb
C_{1,3}$ on 
$V\subset \Bbb C_{1,3}$. One can think of (\ref{L4}) as being the
representation of covectors to $X^4$ by the Dirac $\g$-matrices
on elements of the spinor bundle $S^h$. 
Relative to an atlas $\{z_\zeta\}$ of $P^h$ and to 
the associated Lorentz atlas
$\{z_h\circ z_\zeta\}$ of $LX$, the
representation (\ref{L4}) reads
\be
y^A(\g_h(h^a(x) \ot v))=\g^{aA}{}_By^B(v), \qquad v\in S^h_x,
\ee
where $y^A$ are the
corresponding 
bundle coordinates of $S^h$, and $h^a$ are the tetrad coframes
(\ref{b3211'}). 
For brevity, we will write
\be
\wh h^a=\g_h(h^a)=\g^a,\qquad 
\wh dx^\la=\g_h(dx^\la)=h^\la_a(x)\g^a.
\ee

Let $A_h$ be a principal connection on $S^h$ and let
\be
D: J^1S^h\to T^*X\op\ot_{S^h} S^h,\qquad
D=(y^A_\la-A^{ab}{}_\la L_{ab}{}^A{}_By^B)dx^\la\ot\dr_A,
\ee
be the corresponding covariant differential (\ref{2116}), where
\be
VS^h= S^h\op\times_XS^h.
\ee
The first order
differential  Dirac operator is defined on $S^h$
by the composition 
\mar{l13}\ben
&& \Delta_h=\g_h\circ D: J^1S^h\to T^*X\ot S^h\to S^h, \label{l13}\\
&& y^A\circ\Delta_h=h^\la_a \g^{aA}{}_B(y^B_\la- \frac12 A^{ab}{}_\la
L_{ab}{}^A{}_By^B). \nonumber
\een

Note that there is one-to-one correspondence between the principal
connections, called spin connections, on the
$h$-associated principal spinor bundle
$P^h$ and the Lorentz connections on the L-principal bundle
$L^hX$.
Indeed, it follows from Theorem \ref{mos253} that every principal connection 
\mar{b3205'}\beq
A_h=dx^\la\ot(\dr_\la + \frac12A_\la{}^{ab} \ve_{ab}) 
\label{b3205'}
\eeq
on $P^h$ defines a principal connection on
$L^hX$ which is given by the same expression (\ref{b3205'}).
Conversely, the pull-back $z^*_h\ol A_h$ on
$P^h$ of the connection form $\ol A_h$ of a Lorentz connection
$A_h$ on $L^hX$ is
equivariant under the action of group
$L_\rs$ on
$P^h$ and, consequently, it is a connection form of a spin connection on
$P^h$. 

In particular, the Levi--Civita connection of a pseudo-Riemannian metric $g$
gives rise to the spin connection with the components
\mar{b3217}\beq
A_\la{}^{ab}=\eta^{kb}h^a_\m(\dr_\la h^\m_k - h^\nu_k\{_\la{}^\m{}_\nu\}).
\label{b3217}
\eeq

We consider the general case of a spin connection
generated on $P^h$ by a world connection $K$.
The Lorentz connection $K_h$ induced by $K$ on $L^hX$ takes the
form (\ref{b3205}) with components (\ref{K102}). It defines the
spin connection 
\mar{b3212}\beq
K_h=dx^\la\ot[\dr_\la +\frac14 (\eta^{kb}h^a_\m-\eta^{ka}h^b_\m)(\dr_\la
h^\m_k - h^\nu_k K_\la{}^\m{}_\nu)L_{ab}{}^A{}_B y^B\dr_A] \label{b3212}
\eeq
on $S^h$, where $L_{ab}$ are the generators (\ref{b3213})  
\cite{book00,pon,sard97b}. 
Substituting this spin connection in the Dirac
operator (\ref{l13}), we
obtain the dynamics of Dirac spinor fields in the presence of an arbitrary
world connection on a world manifold, not only of the Lorentz type.

Motivated by the connection (\ref{b3212}), one
can obtain  the lift
\mar{b3216}\beq
\wt\tau = \tau^\la\dr_\la +\frac14
(\eta^{kb}h^a_\m-\eta^{ka}h^b_\m) (\tau^\la\dr_\la
h^\m_k - h^\nu_k\dr_\nu\tau^\m) L_{ab}{}^A{}_B y^B\dr_A \label{b3216}
\eeq
of vector fields $\tau$ on $X$ onto the spinor bundle $S^h$
\cite{book,book00,sard97}. 
This lift can be brought into the form
\be
\wt\tau = \tau_{\{\}} - \frac14
(\eta^{kb}h^a_\m-\eta^{ka}h^b_\m)h^\nu_k\nabla_\nu\tau^\m L_{ab}{}^A{}_B
y^B\dr_A,
\ee
where $\tau_{\{\}}$ is the horizontal lift of $\tau$ by means
of the spin Levi--Civita connection  
for the tetrad field $h$, and
$\nabla_\nu \tau^\m$ are the covariant derivatives of $\tau$ relative to the
Levi--Civita connection \cite{fat98,god,koss}.

Let us further assume that a world manifold $X$ is non-compact, and let
it be parallelizable in order to admit a spin structure. In this case,
all Dirac spin structures are equivalent, i.e., the principal spinor bundles
$P^h$ and $P^{h'}$ are isomorphic \cite{avis,ger}. 
Nevertheless, the
associated structures of the bundles of 
Minkowski spaces $M^hX$ and $M^{h'}X$ (\ref{b3192}) on the cotangent
bundle $T^*X$ are not equivalent, and so are the
representations
$\g_h$ and $\g_{h'}$ (\ref{L4}) \cite{book,sardz92}.
It follows that every Dirac fermion field must be described in a pair
$(s_h,h)$ with a certain tetrad field $h$, and Dirac fermion fields in
the presence of different tetrad fields fail to be given by sections of
the same spinor bundle. The goal is to construct a bundle over $X$
whose sections exhaust the whole totality of fermion-gravitation pairs
\cite{book,book00,sard98a}. Following the general scheme of
describing symmetry breaking in Section 3, we will use the fact that 
the frame bundle $LX$ is the principal bundle $LX\to\Si$ over the
tetrad bundle $\Si$ (\ref{5.15}) with the structure Lorentz
group L.

Let us consider the above mentioned 
two-fold covering bundle $\wt{LX}$ of the
frame bundle $LX$ and the Dirac spin structure (\ref{b3250}) on the
tetrad bundle $\Si$.
Owing to the commutative diagram (\ref{b3243}), we have the commutative
diagram
\mar{b3222}\beq
\begin{array}{ccc}
 \wt{LX} & \ar^{\wt z} &  LX \\
 \put(0,-10){\vector(0,1){20}} & 
& \put(0,-10){\vector(0,1){20}}  \\
P^h & \ar^{z_h} & L^hX 
\end{array} \label{b3222}
\eeq 
for any tetrad field $h$ \cite{fulp,book00,sard98a}. This means
that, given a tetrad field $h$, the restriction $h^*\wt{LX}$ of the
$L_\rs$-principal bundle (\ref{mos265}) to $h(X)\subset
\Si$ is isomorphic to the $L_\rs$-principal subbundle $P^h$ of
the fibre bundle $\wt{LX}\to X$ which is the Dirac spin structure
associated with the Lorentz structure $L^hX$.

Let us consider the spinor bundle
\mar{yy1}\beq
S=(\wt{LX}\times V)/L_\rs\to\Si, \label{yy1}
\eeq
associated with the $L_\rs$-principal bundle (\ref{mos265}),
and the corresponding composite spinor bundle (\ref{y2}).
Given a tetrad field $h$, there is the above mentioned canonical isomorphism 
\be
i_h: S^h=(P^h\times V)/L_\rs \to (h^*\wt{LX}\times V)/L_\rs
\ee
of the $h$-associated spinor bundle $S^h$ (\ref{y1}) onto the
restriction $h^*S$ of the spinor bundle $S\to \Si$ to $h(X)\subset
\Si$ (see Proposition \ref{comp10}). Thence, every global section $s_h$ of the spinor bundle $S^h$
corresponds to the global section $i_h\circ s_h$ of the composite spinor 
bundle (\ref{y2}). Conversely, every global section $s$ of the composite
spinor bundle (\ref{y2}), which projects onto a tetrad field
$h$, takes its values into the
subbundle $i_h(S^h)\subset S$ (see Proposition \ref{mos61}).

Let the frame bundle $LX\to X$ be provided with a holonomic atlas
$\{U_\zeta, T\f_\zeta\}$, 
and let the principal bundles $\wt{LX}\to \Si$ and
$LX\to\Si$ have the associated atlases $\{U_\e,z^s_\e\}$ and
$\{U_\e,z_\e=\wt z\circ z^s_\e\}$, respectively. 
With these atlases, the composite spinor
bundle
$S$ (\ref{y2}) is equipped with the bundle coordinates
$(x^\la,\si_a^\m, y^A)$, where $(x^\la,
\si_a^\m)$ are coordinates of the tetrad bundle $\Si$ such that
$\si^\m_a$ are the matrix components of the group element
$(T\f_\zeta\circ z_\e)(\si),$
$\si\in U_\e,\, \pi_{\Si X}(\si)\in U_\zeta.$
For any tetrad field $h$, we have
$(\si^\la_a\circ h)(x)= h^\la_a(x)$ where $h^\la_a(x)=H^\la_a\circ
z_\e\circ h$ are the tetrad
functions (\ref{L6}) with respect to the Lorentz atlas $\{z_\e\circ
h\}$ of $L^hX$. 

The spinor bundle $S\to\Si$ is
a subbundle of the bundle of Clifford algebras which is generated by 
the bundle of Minkowski spaces  
\mar{mos266}\beq
E_M=(LX\times M)/\rL\to\Si \label{mos266}
\eeq
associated with the L-principal bundle
$LX\to\Si$. Since the fibre bundles $LX\to X$ and 
$GL_4\to GL_4/$L are trivial, so is the fibre
bundle (\ref{mos266}). Hence, it is
isomorphic to the product $\Si\op\times_X T^*X$. 
Then there exists the representation 
\mar{L7}\beq
\g_\Si: T^*X\op\ot_{\Si} S= (\wt{LX}\times (M\ot V))/L_\rs
\to (\wt{LX}\times\g(M\ot V))/L_\rs=S, \label{L7}
\eeq
given by the coordinate expression
\be
\wh dx^\la=\g_\Si (dx^\la) =\si^\la_a\g^a.
\ee
Restricted to $h(X)\subset \Si$, this representation  recovers the
morphism
$\g_h$ (\ref{L4}).

Using the representation $\g_\Si$ (\ref{L7}), one can construct the total Dirac
operator on the composite spinor bundle $S$ as follows. 
Since the bundles $\wt{LX}\to \Si$ and $\Si\to X$ 
are trivial,
let us consider a principal connection $A$ (\ref{b3228}) on the
$L_\rs$-principal bundle $\wt{LX}\to\Si$ given
by the local connection form
\mar{L10,M4}\ben
&& A = (A_\la{}^{ab} dx^\la+ A^k_\m{}^{ab} d\si^\m_k)\ot L_{ab},
\label{L10}\\
&& A_\la{}^{ab} =-\frac12 (\eta^{kb}\si^a_\m-\eta^{ka}\si^b_\m)
 \si^\nu_k K_\la{}^\m{}_\nu, \nonumber\\
&& A^k_\m{}^{ab}=\frac12(\eta^{kb}\si^a_\m -\eta^{ka}\si^b_\m), \label{M4}
\een
where $K$ is a world connection on $X$. 
This connection  defines the associated spin connection
\mar{b3266}\beq
A_\Si = dx^\la\ot(\dr_\la + \frac12A_\la{}^{ab}L_{ab}{}^A{}_By^B\dr_A) +
d\si^\m_k\ot(\dr^k_\m +  \frac12A^k_\m{}^{ab}L_{ab}{}^A{}_By^B\dr_A)
\label{b3266}
\eeq
on the spinor bundle $S\to\Si$. The choice of the connection
(\ref{L10}) is motivated by the fact that, given a tetrad field $h$,
the restriction of the spin
connection (\ref{b3266}) to $S^h$ is exactly the spin connection
(\ref{b3212}).

The connection (\ref{b3266}) yields the first order differential
operator $\wt D$ (\ref{7.10}) on the composite spinor bundle $S\to X$ which
reads
\mar{7.10'}\ben 
&&\wt D:J^1S\to T^*X\op\ot_{\Si} S,\nonumber\\
&&\wt D=dx^\la\ot[y^A_\la- \frac12(A_\la{}^{ab} + A^k_\m{}^{ab}\si_{\la
k}^\m)L_{ab}{}^A{}_By^B]\dr_A  =\label{7.10'} \\
&& \qquad dx^\la\ot[y^A_\la-
\frac14(\eta^{kb}\si^a_\m -\eta^{ka}\si^b_\m)(\si^\m_{\la k} -\si^\nu_k
K_\la{}^\m{}_\nu)L_{ab}{}^A{}_By^B]\dr_A. \nonumber
\een
The corresponding restriction $\wt D_h$ (\ref{b3260}) of the
operator $\wt D$ (\ref{7.10'}) to
$J^1S^h\subset J^1S$ recovers the familiar covariant differential on the
$h$-associated spinor bundle $S^h\to X^4$ relative to the spin connection
(\ref{b3216}).

Combining (\ref{L7}) and (\ref{7.10'}) gives 
the first order differential operator
\mar{3261}\ben
&& \Delta=\g_{\Si}\circ\wt D:J^1S\to T^*X\op\ot_\Si
S\to S, \label{b3261}\\
&& y^B\circ\Delta=\si^\la_a\g^{aB}{}_A[y^A_\la-
\frac14(\eta^{kb}\si^a_\m -\eta^{ka}\si^b_\m)(\si^\m_{\la k} -\si^\nu_k
K_\la{}^\m{}_\nu)L_{ab}{}^A{}_By^B], \nonumber
\een
on the composite spinor bundle $S\to X$.
One can think of $\Delta$ as being the  total Dirac operator  on
$S$ since, for every tetrad field $h$, the restriction of $\Delta$ to
$J^1S^h\subset J^1S$  is exactly the 
Dirac operator $\Delta_h$ (\ref{l13}) on
the spinor bundle
$S^h$
in the presence of the background tetrad field $h$ and the spin connection
(\ref{b3212}).

Thus, we come to the gauge model of metric-affine
gravity and Dirac fermion fields.
The total configuration space of this model is the jet manifold
$J^1Y$ of the bundle product
$Y=C_K\op\times_\Si S$,
where 
$C_K$ is the bundle of world connections (\ref{015}). This
configuration space is 
coordinated by $(x^\m,\si^\m_a, k_\m{}^\al{}_\bt,y^A)$.

Let $J^1_\Si Y$ denote the first order jet manifold of the
fibre bundle $Y\to \Si$. This fibre bundle can be endowed with
the spin connection
\mar{b3263}\ben
&& A_Y:Y\ar J^1_\Si Y\ar^{{\rm pr}_2} J^1_\Si S,\nonumber
\\
&& A_Y = dx^\la\ot(\dr_\la +\wt A_\la{}^{ab}L_{ab}{}^A{}_By^B\dr_A) +
d\si^\m_k\ot(\dr^k_\m +  A^k_\m{}^{ab}L_{ab}{}^A{}_By^B\dr_A), \label{b3263}
\een
where $A^k_\m{}^{ab}$ is given by the expression
(\ref{M4}) and
\be
\wt A_\la{}^{ab} =-\frac12 (\eta^{kb}\si^a_\m-\eta^{ka}\si^b_\m)
 \si^\nu_k k_\la{}^\m{}_\nu.
\ee
Using the connection (\ref{b3263}), we obtain the first order
differential operator
\mar{7.100}\ben 
&&\wt D_Y:J^1Y\to T^*X\op\ot_\Si S,\nonumber\\
&&\wt D_Y=dx^\la\ot[y^A_\la- 
\frac14(\eta^{kb}\si^a_\m -\eta^{ka}\si^b_\m)(\si^\m_{\la k} -\si^\nu_k
k_\la{}^\m{}_\nu)L_{ab}{}^A{}_By^B]\dr_A, \label{7.100}
\een
and the total Dirac operator
\mar{b3264}\ben
&& \Delta_Y=\g_\Si\circ\wt D:J^1Y\to T^*X\op\ot_\Si S\to S,
\nonumber\\
&& y^B\circ\Delta_Y=\si^\la_a\g^{aB}{}_A[y^A_\la-  \frac14(\eta^{kb}\si^a_\m
-\eta^{ka}\si^b_\m)(\si^\m_{\la k} -\si^\nu_k
k_\la{}^\m{}_\nu)L_{ab}{}^A{}_By^B], \label{b3264}
\een
on  the fibre bundle $Y\to X$. 
Given a world connection 
$K:X\to C_K$, the restrictions of the spin connection $A_Y$ (\ref{b3263}),
the operator $\wt D_Y$ (\ref{7.100}) and the Dirac operator $\Delta_Y$
(\ref{b3264}) to $K^*Y$ are exactly the 
spin connection (\ref{b3266}) and the
operators (\ref{7.10'}) and (\ref{b3261}), respectively.

Finally, since the spin structure (\ref{b3250})
is unique, the
$\wt{GL}_4$-principal bundle
$\wt{LX}\to X^4$ as well as the frame bundle $LX$ admits the canonical
lift of any diffeomorphism $f$ of the base $X$ \cite{dabr,henn,book00}.
This lift yields the general covariant transformation of the associated
spinor bundle
$S\to\Si$ over the general covariant transformations of the
tetrad bundle $\Si$.                 
The corresponding canonical lift onto $S$ of a vector field on $X$ can be
constructed as a generalization of the lift (\ref{b3216}) (see
\cite{book,sard98a} for detail). 

\section{Appendix. Fibre bundles and jet manifolds}

All maps throughout are smooth and manifolds are real,
finite-dimensional, Hausdorff, second-countable (hence, paracompact)
and connected.

\glos{Fibre bundles}

A manifold $Y$ is called a fibred manifold over a base $X$ if there is
a surjective submersion
\mar{z10}\beq
\pi:Y\to X, \label{z10}
\eeq
i.e., the tangent map $T\pi:TY\to TX$ is a surjection.
A fibred manifold $Y\to X$ is
provided with an atlas of fibred coordinates $(x^\la,y^i)$, where
$x^\la$ are coordinates on the base $X$, whose transition functions
$x^\la\to x'^\la$ are
independent of the coordinates $y^i$.

A fibred manifold $Y\to X$ is called a fibre bundle if it is locally trivial. 
This means that
the base
$X$ admits an open covering 
$\{U_\xi\}$ so that $Y$ is 
locally equivalent to the splittings 
\mar{mos02}\beq
\psi_\xi:\pi^{-1}(U_\xi) \to U_\xi\times V \label{mos02}
\eeq
together with the 
transition functions 
\mar{mos271}\beq
\psi_\xi (y)=(\rho_{\xi\zeta}\circ\psi_\zeta)(y),
\qquad y\in \pi^{-1}(U_\xi\cap U_\zeta). \label{mos271}
\eeq
The manifold $V$ is one for all local splittings (\ref{mos02}). It is
called the typical fibre
of $Y$. Local
trivializations  
$(U_\xi, \psi_\xi)$ make up a bundle atlas
$\Psi = \{(U_\xi, \psi_\xi)\}$
of the fibre bundle $Y$.
Given an atlas $\Psi$, a fibre bundle $Y$ is provided with the
associated bundle coordinates $(x^\la,y^i)$ where
\be
y^i(y)=(y^i\circ\pr_2\circ\psi_\xi)(y), \qquad y\in Y,
\ee
are coordinates on the typical fibre $V$. Note that a fibre bundle $Y\to X$
is uniquely defined by a bundle atlas $\Psi$. Two bundle atlases are
said to be equivalent if a union of these atlases is also a bundle
atlas.

A fibre bundle $Y\to X$ is called trivial if $Y$ is diffeomorphic to
the product $X\times V$. Any fibre
bundle over a contractible base is trivial. 

By a (global) section of a fibre bundle (\ref{z10}) is meant a manifold
morphism 
$s:X\to Y$ such that $\pi\circ s=\id X$. A section $s$ is
an imbedding, i.e., $s(X)\subset Y$ is both a
(closed) submanifold and a 
topological subspace of $Y$.
Similarly, a section  $s$ of a fibre bundle $Y\to
X$ over a submanifold $N\subset X$ is a morphism  
$s:N\to Y$ such that $\pi\circ s:N\hookrightarrow X$
is a natural injection.
 A section of a fibre bundle over an
open subset $U\subset X$ is simply called a local
section. A fibre bundle, by definition, admits a local section 
around each point of its base. 

\begin{theo} \label{mos9} \mar{mos9} 
Any fibre bundle $Y\to X$ whose typical fibre is diffeomorphic to $\Bbb
R^m$ has a global section. Its section over a closed
subset of $N\subset X$ can always be extended to a global section. 
\end{theo}

A bundle morphism of two bundles
$\pi:Y\to X$ and $\pi': Y'\to X'$ is a pair of maps $\Phi:Y\to Y'$ and
$f:X\to X'$ such that the diagram
\mar{r16}\beq
\begin{array}{rcccl}
& {Y} &  \op\longrightarrow^{\Phi} & {Y'} &  \\
{_\pi} &\put(0,10){\vector(0,-1){20}} & & \put(0,10){\vector(0,-1){20}} &
{_{\pi'}}
\\ & {X} & \op\longrightarrow_{f} & {X'} &
\end{array} \label{r16}
\eeq
is commutative, i.e., $\Phi$ sends fibres to fibres. In brief, one says
that (\ref{r16}) is a bundle morphism $\Phi$
over $f$. If $f=\id X$, then $\Phi$
is called a bundle morphism over $X$.

If a bundle morphism $\Phi$ (\ref{r16})
is a diffeomorphism, it is called an isomorphism
of fibre bundles. Two fibre bundles over the same
base $X$ are said to be
equivalent if there exists their
isomorphism over $X$. 
A bundle morphism $\Phi$ (\ref{r16}) over $X$ (or its image $\Phi(Y)$)
is called a subbundle of the fibre bundle $Y'\to X$ if
$\Phi(Y)$ is a submanifold of $Y'$. 

Given a fibre bundle $\pi:Y\to X$ and a morphism $f: X'\to X$, the
pull-back of $Y$ by $f$ is the manifold
\mar{mos106}\beq
f^*Y =\{(x',y)\in X'\times Y \,: \,\, \pi(y) =f(x')\} \label{mos106}
\eeq
together with the natural projection $(x',y)\mapsto x'$. It is a fibre
bundle over $X'$ such that the fibre of $f^*Y$
over a point
$x'\in X'$ is that of
$Y$ over the point $f(x')\in X$. 

In particular, if $i_{X'}:X'\subset X$
is a submanifold, then the pull-back 
$i_{X'}^*Y=Y\mid_{X'}$ is called the restriction
of a fibre bundle $Y$ to the submanifold $X'\subset X$.

Let $\pi:Y\to X$ and $\pi':Y'\to X$ be fibre bundles
over the same base 
$X$. Their fibred product $Y\op\times Y'$
over $X$ is defined as the pull-back 
\be
Y\op\times_X Y'=\pi^*Y'\quad {\rm or} \quad 
Y\op\times_X Y'={\pi'}^*Y.
\ee

\glos{Vector and affine bundles} 

A vector bundle is a fibre bundle 
$Y\to X$ such that its
typical fibre $V$ and all fibres $Y_x=\pi^{-1}(x)$, $x\in X$, are
real finite-dimensional vector spaces, and
there is a linear bundle atlas $\Psi=\{(U_\xi,\psi_\xi)\}$ of $Y\to X$
whose trivialization morphisms 
$\psi_\xi(x):Y_x\to V$, $x\in U_\xi$ are linear isomorphisms.
The associated bundle
coordinates $(y^i)$ have linear transition functions.
By a morphism of vector bundles is meant a bundle morphism 
whose restriction to each fibre  
of $Y$ is a linear map. 

There are the following standard constructions of new vector bundles from
old ones.
\begin{itemize}
\item Let $Y\to X$ be a vector bundle with a typical fibre $V$. By
$Y^*\to X$ is meant the dual vector bundle
with the typical fibre $V^*$ dual of $V$.
The interior product
of $Y$ and $Y^*$ is defined as a bundle morphism 
\be
\rfloor: Y\otimes Y^*\ar_X X\times \Bbb R.
\ee
\item Let $Y\to X$ and $Y'\to X$ be vector bundles with typical fibres
$V$ and $V'$, respectively. Their Whitney sum 
$Y\op\oplus_X Y'$ is a vector bundle over $X$ with the typical fibre 
$V\oplus V'$.
\item Let $Y\to X$ and $Y'\to X$ be vector bundles with typical fibres
$V$ and $V'$, respectively. Their tensor product 
$Y\op\ot_X Y'$ is a vector bundle over $X$ 
with the typical fibre $V\ot V'$. Similarly, the exterior product
of vector bundles $Y\op\w_X Y$  is defined.  
\end{itemize}

By virtue of Theorem \ref{mos9}, vector bundles have global
sections. Moreover, there exists the canonical zero-valued global section $\wh
0$ of vector bundles.

Let us consider an exact sequence of vector
bundles over the same base $X$ 
\mar{ms0100}\beq
0\to Y'\op\to^i Y\op\to^j Y''\to 0, \label{ms0100}
\eeq
where $Y'\to Y$ is an injection
and $Y\to Y''$ is a surjection of vector bundles such that $\im i =\Ker j$. 
This is equivalent to the fact that $Y''=Y/Y'$ is the quotient bundle. 
One says that the exact sequence (\ref{ms0100}) of
vector bundles admits a splitting
if there exists a linear bundle monomorphism
$\G:Y''\to Y$ over $X$  such that $j\circ\G=\id Y''$. Then
\be
Y=i(Y')\oplus \G(Y'').
\ee
Every exact sequence of vector bundles admits a splitting.

Let $\ol\pi:\ol Y\to X$ be a vector bundle with a typical fibre $\ol V$.
An affine bundle modelled over the vector
bundle $\ol Y\to X$ is a fibre bundle $\pi:Y\to X$ whose
typical fibre $V$ is an affine space modelled over $\ol V$, while the
following conditions hold.
(i) All the fibres $Y_x$ of $Y$ are affine spaces modelled over the
corresponding fibres $\ol Y_x$ of the vector bundle $\ol Y$.
(ii) There is a bundle atlas $\Psi=\{(U_\al,\psi_\al)\}$ of $Y\to X$
whose trivializations morphisms
$\psi_\zeta(x):Y_x\to V$, $x\in U_\zeta$, are affine isomorphisms.
The associated bundle coordinates $(y^i)$ have affine transition functions.
There are the bundle morphisms
\be
&&Y\op\times_X\ol Y\ar_X Y,\qquad (y^i, \ol y^i)\mapsto  y^i +\ol y^i,\\
&&Y\op\times_X Y\ar_X \ol Y,\qquad (y^i, y'^i)\mapsto  y^i - y'^i,
\ee
where $(\ol y^i)$ are linear coordinates on the vector bundle
$\ol Y$.  

In particular, every vector bundle has a natural structure of an affine
bundle.

By a morphism of affine bundles
is meant a bundle morphism 
whose restriction to each fibre  
of $Y$ is an affine map. 
Every affine bundle morphism $\Phi:Y\to Y'$ from an affine bundle $Y$ modelled
over a vector bundle $\ol Y$ to an affine bundle
$Y'$ modelled over a vector bundle $\ol Y'$
 determines uniquely the linear bundle morphism  
\mar{1355'}\beq
\ol \Phi: \ol Y\to \ol Y', \qquad
\ol y'^i\circ \ol\Phi= \frac{\dr\Phi^i}{\dr y^j}\ol y^j, \label{1355'}
\eeq
called the linear derivative of $\Phi$.

By virtue of Theorem \ref{mos9}, affine bundles  have  global sections.
Let $\pi:Y\to X$ be an affine bundle modelled over a vector bundle $\ol
Y\to X$. Every global section $s$ of an affine bundle $Y\to X$ yields
the bundle morphism
\be
D_s:Y\ni y\to y-s(\pi(y))\in \ol Y. 
\ee

\glos{Tangent and cotangent bundles} 

Tangent and cotangent bundles exemplify vector bundles.
The fibres of the tangent bundle $\pi_Z:TZ\to Z$
of a manifold $Z$ are tangent spaces to $Z$.
Given an atlas $\Psi_Z =\{(U_\xi,\phi_\xi)\}$
of a manifold $Z$, the tangent bundle is provided with the holonomic
atlas
\mar{mos150}\beq
\Psi =\{U_\xi, \psi_\xi = T\phi_\xi)\}, \label{mos150}
\eeq
where by $T\phi_\xi$ is meant the tangent map to $\f_\xi$.
The associated linear bundle coordinates  are coordinates
with respect to
the  holonomic frames  $\{\dr_\la\}$ in tangent spaces
$T_zZ$. They are called  
holonomic coordinates, and are denoted   
by $(\dot
z^\la)$ on $TZ$. The transition functions of holonomic coordinates read 
\be
\dot z'^\la=\frac{\dr z'^\la}{\dr z^\m}\dot z^\m. 
\ee

Every manifold map $f:Z\to Z'$ generates the linear bundle morphism
of the tangent bundles 
\be
Tf: TZ\ar_f TZ',\qquad \dot z'^\la\circ Tf=\frac{\dr f^\la}{\dr z^\m}\dot z^\m,
\ee
called the tangent map to $f$.

The cotangent bundle of a manifold $Z$
is the dual $\pi_{*Z}:T^*Z\to Z$
of the tangent bundle $TZ\to Z$. It is equipped with the holonomic
coordinates $(z^\la,\dot z_\la)$ with respect to the 
coframes   $\{dz^\la\}$ dual of
$\{\dr_\la\}$. Their transition functions read
\be
\dot z'_\la=\frac{\dr z^\m}{\dr z'^\la}\dot z_\m.
\ee
Various tensor products
\mar{mos6}\beq
T=(\op\ot^mTZ)\ot(\op\ot^kT^*Z) \label{mos6}
\eeq
over $Z$ of tangent and cotangent bundles are called tensor bundles.

Let $\pi_Y:TY\to Y$ be the tangent bundle of a fibre bundle
$\pi: Y\to X$.
Given bundle coordinates $(x^\la,y^i)$ on $Y$, the tangent bundle $TY$
is equipped with the holonomic coordinates 
$(x^\la,y^i,\dot x^\la, \dot y^i)$.
The tangent bundle $TY\to Y$ has the
vertical tangent subbundle $VY = \Ker T\pi$,
given by the coordinate relations $\dot x^\la =0$. This subbundle
consists of the vectors tangent to fibres of $Y$. The vertical tangent bundle
$VY$ is provided with the holonomic coordinates $(x^\la,y^i,\dot y^i)$ with
respect to the  frames $\{\dr_i\}$.

Let $T\Phi$ be the tangent map to a fibred morphism $\Phi:Y\to Y'$.
Its restriction 
\be
V\Phi=T\Phi |_{VY}:VY\to VY',\qquad \dot y'^i\circ V\Phi
=\dr_V\Phi^i=\dot y^j\dr_j\Phi^i,  
\ee
to $VY$ is a linear bundle morphism of the vertical tangent bundle $VY$ to
the vertical tangent bundle $VY'$, called the vertical tangent map
to $\Phi$.

Vertical tangent bundles of many fibre bundles are equivalent
to the pull-backs
\be
VY=Y\op\times_X\ol Y
\ee
where $\ol Y\to X$ is some vector bundle. This means that the transition
functions of the holonomic coordinates $\dot y^i$ on $VY$ are independent of
$y^i$. In particular, every affine bundle $Y\to X$ modelled over a 
vector bundle $\ol Y\to X$
admits the canonical vertical splitting 
\mar{48}\beq
VY\cong Y\op\times_X\overline Y\label{48}
\eeq
because the  coordinates $\dot y^i$ on $VY$ have the same
transformation law as the linear coordinates $\ol y^i$ on the vector
bundle $\ol Y$. If $Y$ is a vector bundle, the splitting (\ref{48})
reads
\mar{1.10}\beq
VY\cong Y\op\times_X Y. \label{1.10}
\eeq

The vertical cotangent bundle $V^*Y\to Y$ of a fibre bundle $Y\to X$ 
is defined as the vector bundle dual
of the vertical tangent bundle $VY\to Y$. There is
the canonical projection 
\mar{z11}\beq
\zeta: T^*Y\op\to_YV^*Y, \qquad
\zeta: \dot x_\la dx^\la +\dot y_i dy^i \mapsto \dot y_i \ol
dy^i,\label{z11} 
\eeq
where $\{\ol dy^i\}$ are the  bases for the fibres of $V^*Y$, which
are dual of the holonomic frames $\{\dr_i\}$ for the vertical tangent
bundle $VY$.

With $VY$ and $V^*Y$, we have the following two exact sequences of
vector bundles over $Y$:
\mar{1.8ab}\bea
&& 0\to VY\hookrightarrow TY\op\to^{\pi_T} Y\op\times_X TX\to 0,
\label{1.8a} \\
&& 0\to Y\op\times_X T^*X\hookrightarrow T^*Y\op\to^\zeta V^*Y\to 0.
\label{1.8b}
\eea
Their splitting
corresponds to the choice of a connection on the
fibre bundle $Y\to X$.

\glos{Vector fields}

A vector field on a manifold $Z$ is defined
as a global section of the tangent bundle $TZ\to Z$.  
The set $\cT(Z)$ of vector fields on $Z$ is a real Lie
algebra with respect to the Lie bracket
\be
[v,u] = (v^\la\dr_\la u^\m - u^\la\dr_\la v^\m)\dr_\m, \quad u=u^\la\dr_\la,
\quad v=v^\la\dr_\la. 
\ee
Every vector field on a manifold $Z$ is the generator
of a local 1-parameter group of local diffeomorphisms of $Z$ \cite{kob}.
A vector field $u$ on a manifold $Z$ is called complete if it is
induced by a 1-parameter group of 
diffeomorphisms of $Z$.

A vector field $u$
on a fibre bundle $Y\to X$ is said to be projectable
if it projects over a vector field 
$\tau$ on $X$, i.e., $\tau\circ\pi= T\pi\circ u$. 
It has the coordinate expression
\be
u=u^\la(x^\m) \dr_\la + u^i(x^\m,y^j) \dr_i, \qquad \tau=u^\la\dr_\la.
\ee
A projectable vector field $u=u^i\dr_i$ on a fibre bundle $Y\to X$ 
is said to be vertical if it
projects over the zero vector field $\tau=0$ on $X$.  

A vector field $\tau=\tau^\la\dr_\la$ on a base $X$ of a
fibre bundle $Y\to X$ gives rise to a projectable vector field on the
total space $Y$ by means of some connection on this fibre bundle. 
Nevertheless, every tensor bundle (\ref{mos6})
admits the canonical lift
\mar{l28}\beq
\wt\tau = \tau^\m\dr_\m + [\dr_\nu\tau^{\al_1}\dot
x^{\nu\al_2\cdots\al_m}_{\bt_1\cdots\bt_k} + \ldots
-\dr_{\bt_1}\tau^\nu \dot x^{\al_1\cdots\al_m}_{\nu\bt_2\cdots\bt_k}
-\ldots]\frac{\dr}{\dr \dot
x^{\al_1\cdots\al_m}_{\bt_1\cdots\bt_k}} \label{l28}
\eeq
of any vector field $\tau$ on $X$. In particular, we have the
canonical lift  
\mar{l27}\beq
\wt\tau = \tau^\m\dr_\m +\dr_\nu\tau^\al\dot x^\nu\dot\dr_\al, \qquad 
\dot\dr_\la = \frac{\dr}{\dr\dot x^\la}, \label{l27}
\eeq
of $\tau$ onto the tangent bundle $TX$.

\glos{Exterior forms} 

An exterior $r$-form on a manifold $Z$ is a
section 
\be
\f =\frac{1}{r!}\f_{\la_1\dots\la_r}
dz^{\la_1}\w\cdots\w dz^{\la_r}
\ee
of the exterior product $\op\w^r T^*Z\to Z$. All 
exterior forms on $Z$ constitute the exterior ${\Bbb Z}$-graded
algebra $\cO^*(Z)$  with respect to the exterior
product. It is provided with exterior differential 
\be
d: \cO^r(Z) \to \cO^{r+1}(Z), \qquad 
d\f= \frac{1}{r!}
\dr_\m\f_{\la_1\ldots\la_r} dz^\m\w dz^{\la_1}\w\cdots dz^{\la_r},
\ee
which obeys the relations 
\be
d(\f\w\si)= d(\f)\w \si +(-1)^{\nm\f}\f\w d(\si), \qquad  d\circ d=0.
\ee

Given a morphism $f:Z\to Z'$, by $f^*\f$ is meant the 
pull-back on $Z$ of an $r$-form $\f$ on $Z'$ by $f$,
which is defined by the condition
\be
f^*\f(v^1,\ldots,v^r)(z) = \f(Tf(v^1),\ldots,Tf(v^r))(f(z)), \qquad \forall
v^1,\cdots v^r\in T_zZ,
\ee
and obeys the relations
\be
f^*(\f\w\si) =f^*\f\w f^*\si, \qquad 
 df^*\f =f^*(d\f). 
\ee

Let $\pi:Y\to X$ be a fibre bundle with bundle coordinates $(x^\la,y^i)$. 
The pull-back on $Y$ of exterior forms on $X$ by $\pi$ provides the
inclusion $\pi^*:\cO^*(X)\to \cO^*(Y)$. Elements of its image
are called basic forms. Exterior forms
\be
\phi : Y\to\op\w^r T^*X, \qquad 
\phi =\frac{1}{r!}\phi_{\la_1\ldots\la_r}dx^{\la_1}\w\cdots\w dx^{\la_r},
\ee
on $Y$ such that $\vt\rfloor\f=0$ for arbitrary vertical
vector field $\vt$ on $Y$ are said to be 
horizontal forms. 

The interior product of a vector field
$u = u^\m\dr_\m$ and an exterior $r$-form $\f$ is given by the coordinate
expression
\be
u\rfloor\f = \op\sum^r_{k=1}\frac{-1^{k-1}}{r!}u^{\la_k}\f_{\la_1
\ldots\la_k\ldots\la_r} dz^{\la_1}\w\cdots\w\wh{dz}^{\la_k}\w\cdots
dz^{\la_r}. 
\ee
It satisfies the relations
\be
 \f(u_1,\ldots,u_r)=u_r\rfloor\cdots u_1\rfloor\f,\qquad
 u\rfloor(\f\w\si)= u\rfloor\f\w\si +(-1)^{\nm\f}\f\w u\rfloor\si.
\ee
The Lie derivative of an exterior form
$\f$ along a vector field $u$ is 
\be
\bL_u\f =u\rfloor d\f +d(u\rfloor\f).
\ee

\glos{Tangent-valued forms} 

A tangent-valued $r$-form on a
manifold $Z$ is a section 
\be
\phi = \frac{1}{r!}\phi_{\la_1\ldots\la_r}^\m dz^{\la_1}\w\cdots\w
dz^{\la_r}\ot\dr_\m
\ee
of the tensor bundle $\op\w^r T^*Z\ot TZ\to Z$.

There is one-to-one correspondence between the tangent-valued 1-forms 
$\f$ on a manifold $Z$ and the linear bundle endomorphisms 
\mar{29b,b'}\ben
&& \wh\f:TZ\to TZ,\qquad
\wh\f: T_zZ\ni v\mapsto v\rfloor\f(z)\in T_zZ, \label{29b} \\
&&\wh\f^*:T^*Z\to T^*Z,\qquad 
\wh\f^*: T_z^*Z\ni v^*\mapsto \f(z)\rfloor v^*\in T_z^*Z. \label{29b'} 
\een
In particular, the canonical tangent-valued 1-form
\mar{b1.51}\beq
\th_Z= dz^\la\ot \dr_\la \label{b1.51}
\eeq
on $Z$ corresponds to the identity morphisms (\ref{29b}) and (\ref{29b'}).

The space $\cO^*(M)\ox \cT(M)$ of 
tangent-valued forms is provided with the Fr\" olicher--Nijenhuis
bracket generalizing the Lie bracket of
vector fields as follows:
\be
&& [\; ,\; ]_{\rm FN}:\cO^r(M)\ox \cT(M)\times \cO^s(M)\ox \cT(M) 
\to\cO^{r+s}(M)\ox \cT(M), \nonumber \\
&& [\phi,\si]_{\rm FN} = \frac{1}{r!s!}(\phi_{\la_1
\dots\la_r}^\nu\dr_\n\si_{\la_{r+1}\dots\la_{r+s}}^\m - \si_{\la_{r+1}
\dots\la_{r+s}}^\nu\dr_\nu\phi_{\la_1\dots\la_r}^\m -\\
&&  r\phi_{\la_1\ldots\la_{r-1}\nu}^\m\dr_{\la_r}\si_{\la_{r+1}
\dots\la_{r+s}}^\nu + s \si_{\nu\la_{r+2}\ldots\la_{r+s}}^\m
\dr_{\la_{r+1}}\phi_{\la_1\ldots\la_r}^\nu)dz^{\la_1}\wedge\cdots
\wedge dz^{\la_{r+s}}\otimes\dr_\m,\\
&& \f\in \cO^r(M)\ox \cT(M), 
\qquad \si\in \cO^s(M)\ox \cT(M).
\ee
There are the relations 
\mar{1150}\ben
&& [\f,\s]_{\rm FN}=(-1)^{\nm\f\nm\s+1}[\s,\f]_{\rm FN}, \label{1150} \\
&& [\f, [\s, \th]_{\rm FN}]_{\rm FN} = [[\f, \s]_{\rm FN}, \th]_{\rm
FN} +(-1)^{\nm\f\nm\s}  [\s, [\f,\th]_{\rm FN}]_{\rm FN}. \label{1150'}
\een \mar{1150'}

Given a tangent-valued form  $\th$, the Nijenhuis
differential on $\cO^*(M)\ot\cT(M)$ 
is defined as the morphism
\be
d_\th : \si\mapsto d_\th\si = [\th,\si]_{\rm FN}, \qquad 
\forall \si\in\cO^*(M)\ot\cT(M). 
\ee
By virtue of (\ref{1150'}), it has the property
\be 
d_\f[\psi,\th]_{\rm FN} = [d_\phi\psi,\th]_{\rm FN}+
(-1)^{\mid\f\mid\mid\psi\mid} [\psi,d_\f\th]_{\rm FN}.
\ee

Let $\Y$ be a fibre bundle. One consider the following classes
of tangent-valued forms on $Y$:
\begin{itemize}
\item  tangent-valued horizontal forms
\be
&& \phi : Y\to\op\w^r T^*X\op\otimes_Y TY,\\
&& \phi =dx^{\la_1}\wedge\dots\wedge dx^{\la_r}\otimes
[\phi_{\la_1\dots\la_r}^\m(y) \dr_\m +\phi_{\la_1\dots\la_r}^i(y) \dr_i],
\ee
\item  vertical-valued horizontal forms 
\be
\phi : Y\to\op\w^r T^*X\op\otimes_Y VY,\qquad
\phi =\phi_{\la_1\dots\la_r}^i(y)dx^{\la_1}\wedge\dots
\wedge dx^{\la_r}\otimes\dr_i,
\ee
\item vertical-valued horizontal 1-forms $\si = \si_\la^i(y)
dx^\la\otimes\dr_i$, called soldering forms,  
\item basic soldering forms $\si = \si_\la^i(x) dx^\la\otimes\dr_i$.
\end{itemize}

The tangent bundle $TX$ is provided with the canonical soldering
form 
\mar{z117'}\beq
\th_J=dx^\la\ot\dot\dr_\la. \label{z117'}.
\eeq
Due to the canonical vertical splitting
\mar{mos163}\beq
VTX=TX\op\times_X TX, \label{mos163}
\eeq
the canonical soldering form (\ref{z117'}) on $TX$ defines
the canonical tangent-valued form $\th_X$ (\ref{b1.51}) on $X$.
By this reason, tangent-valued 1-forms on a manifold
$X$ are also called soldering forms.

\glos{First order jet manifolds} 

Given a fibre bundle $Y\to X$ with bundle coordinates $(x^\la,y^i)$,
let us consider the equivalence classes
$j^1_xs$, $x\in X$, of its sections $s$, which are identified by their values
$s^i(x)$ and the values of their first order derivatives 
$\dr_\mu s^i(x)$ at points $x\in X$. The equivalence class $j^1_xs$
is called the first order jet of sections 
$s$ at the point $x\in X$. The set  $J^1Y$ of first order jets is provided with
a manifold structure with respect to the adapted coordinates
$(x^\la,y^i,y_\la^i)$ such that
\mar{50}\ben
&&(x^\la,y^i,y_\la^i)(j^1_xs)=(x^\la,s^i(x),\dr_\la s^i(x)),\nonumber\\
&&{y'}^i_\la = \frac{\dr x^\m}{\dr{x'}^\la}(\dr_\m 
+y^j_\m\dr_j)y'^i.\label{50}
\een
It is called the jet manifold of 
the fibre bundle $Y\to X$. It has the natural fibrations 
\be
\pi^1:J^1Y\ni j^1_xs\mapsto x\in X, \qquad
\pi^1_0:J^1Y\ni j^1_xs\mapsto s(x)\in Y, 
\ee
where the latter
is an affine bundle modelled over the vector
bundle  
\mar{cc9}\beq
T^*X \op\otimes_Y VY\to Y.\label{cc9}
\eeq

The jet manifold
$J^1Y$ admits the canonical imbedding 
\mar{18}\beq
\la_1:J^1Y\op\hookrightarrow_Y
T^*X \op\otimes_Y TY,\qquad \la_1=dx^\la
\otimes (\dr_\la + y^i_\la \dr_i). \label{18}
\eeq
Therefore, one usually identifies $J^1Y$ with its image
under this morphism, and represents jets
by the tangent-valued form
\be
dx^\la \otimes (\dr_\la + y^i_\la \dr_i). 
\ee

Any bundle morphism $\Phi:Y\to Y'$ over a diffeomorphism $f$
is extended to the morphism of the corresponding jet manifolds
\be
J^1\Phi : J^1Y  \op\to_\Phi J^1Y',\qquad
{y'}^i_\la\circ J^1\Phi=\frac{\dr(f^{-1})^\m}{\dr x'^\la}d_\m\Phi^i.
\ee
It is called the jet prolongation of the morphism $\Phi$.
Each section $s$ of a fibre bundle $Y\to X$ has the jet prolongation
to the section
\be
 (J^1s)(x)\op =^\df j_x^1s, \qquad 
(y^i,y_\la^i)\circ J^1s= (s^i(x),\dr_\la s^i(x)),
\ee
of the jet bundle $J^1Y\to X$. 

\section{Appendix. Connections}

A connection on a fibre
bundle
$Y\to X$ is defined as a linear bundle monomorphism 
\mar{150}\beq
\G: Y\op\times_X TX\to TY, \qquad 
\G: \dot x^\la\dr_\la \mapsto \dot x^\la(\dr_\la+\G^i_\la\dr_i),
\label{150}
\eeq
over $Y$ which splits the
exact sequence (\ref{1.8a}) and provides the horizontal decomposition 
\mar{152}\ben
&& TY=\G(Y\times TX)\op\oplus_Y VY, \label{152} \\
&& \dot x^\la \dr_\la + \dot y^i \dr_i = \dot
x^\la(\dr_\la + \G_\la^i \dr_i) + (\dot y^i - \dot
x^\la\G_\la^i)\dr_i,  \nonumber
\een
of the tangent bundle $TY$. 
A connection always exists. 

The morphism $\G$ (\ref{150}) yields uniquely  the
horizontal tangent-valued 1-form
\mar{154}\beq
\G = dx^\la\otimes (\dr_\la +
\G_\la^i\dr_i) \label{154}
\eeq
on $Y$ such that 
\be
\G:\dr_\la\mapsto \dr_\la\rfloor\G=\dr_\la +\G^i_\la\dr_i.
\ee
One can think of it as
being another definition of a connection on a fibre bundle $Y\to X$.

Given a connection $\G$ on a fibre bundle $Y\to X$, every 
vector field $\tau$ on its base $X$ gives rise to the projectable
vector field  
\mar{b1.85}\beq
\G \tau=\tau\rfloor\G=\tau^\la(\dr_\la +\G^i_\la\dr_i) \label{b1.85}
\eeq
on $Y$, called the horizontal
lift of
$\tau$ by the connection $\G$.

\glos{Connections as sections of the jet bundle}

Let $Y\to X$ be a fibre bundle, and $J^1Y$ its first order jet manifold. Given
the canonical morphism (\ref{18}), we have the
corresponding morphism
\mar{z20}\beq
\wh\la_1: J^1Y\op\times_X TX\ni\dr_\la\mapsto d_\la = \dr_\la\rfloor
\la\in J^1Y\op\times_Y TY. \label{z20}
\eeq
This morphism yields the canonical
horizontal splitting of the pull-back
\mar{1.20}\ben
&&J^1Y\op\times_Y TY=\wh\la_1(TX)\op\oplus_{J^1Y} VY,\label{1.20}\\
&& \dot x^\la\dr_\la
+\dot y^i\dr_i =\dot x^\la(\dr_\la +y^i_\la\dr_i) + (\dot y^i-\dot x^\la
y^i_\la)\dr_i. \nonumber
\een
Let $\G$ be a global section of $J^1Y\to Y$. Substituting
the tangent-valued form
\be
\la_1\circ \G= dx^\la\ot(\dr_\la +\G^i_\la\dr_i) 
\ee
in the canonical splitting (\ref{1.20}), we obtain the familiar
horizontal splitting (\ref{152}) of $TY$ by
means of a connection
$\G$ on $Y\to X$. Then one can show that there is one-to-one
correspondence between 
the connections on a fibre bundle 
$Y\to X$ and the global sections 
of the affine jet bundle $J^1Y\to Y$.

It follows that 
connections on a fibre bundle $Y\to X$  make up an affine space modelled over
the vector space of soldering forms on
$Y\to X$, i.e., sections of the vector bundle (\ref{cc9}).
 One
deduces immediately from (\ref{50}) the coordinate transformation law 
\be
\G'^i_\la = \frac{\dr x^\m}{\dr{x'}^\la}(\dr_\m 
+\G^j_\m\dr_j)y'^i.
\ee

Every connection $\G$ on a fibre bundle $Y\to X$ yields
the first order 
differential operator 
\mar{2116}\beq
D_\G:J^1Y\op\to_Y T^*X\op\otimes_Y VY, \qquad
D_\G=\la_1- \G\circ \pi^1_0 =(y^i_\la -\G^i_\la)dx^\la\otimes\dr_i,
\label{2116}
\eeq
called the covariant differential  relative to the connection
$\G$. 
If $s:X\to Y$ is a section, one obtains from (\ref{2116}) its 
covariant differential
\mar{+190}\ben
&&\nabla^\G s  =  D_\G\circ J^1s:X\to T^*X\ox VY, \label{+190}\\
&&\nabla^\G s  =  (\dr_\la s^i - \G_\la^i\circ s) dx^\la\ox \dr_i,
\nonumber
\een
and the covariant derivatives $\nabla_\tau^\G
=\tau\rfloor\nabla^\G$ along a vector field $\tau$ on $X$.
A section $s$ is said to be an 
integral section of a connection $\G$ if
$\nabla^\G s=0$. 

Let us recall the following standard constructions of new connections
from the old ones.

Given a fibre bundle $Y\to X$, let $f:X'\to X$ be a map and $f^*Y\to X'$
the pull-back of $Y$ by $f$. Any connection $\G$ 
on $Y\to X$ induces the
pull-back connection
\mar{mos82}\beq
f^*\G=(dy^i-(\G\circ f_Y)^i_\la\frac{\dr f^\la}{\dr x'^\m}dx'^\m)\ot\dr_i
\label{mos82}
\eeq
on $f^*Y\to X'$.

Let $Y$ and $Y'$ be fibre bundles over the same base $X$.
Given a connection $\G$ on $\Y$ and a connection $\G'$ on $Y'\to
X$, the fibre bundle $Y\op\times_X Y'\to X$  is provided with the 
product connection  
\mar{b1.96}\ben
&&\G\times\G': Y\op\times_XY'\to
J^1(Y\op\times_XY')=J^1Y\op\times_XJ^1Y', \nonumber\\
&&\G\times\G' = dx^\la\ot(\dr_\la +\G^i_\la\frac{\dr}{\dr y^i} +
\G'^j_\la\frac{\dr}{\dr y'^j}). \label{b1.96}
\een

Let $i_Y:Y\to Y'$ be a subbundle of a fibre bundle  $Y'\to X$ and
$\G'$ a connection on $Y'\to X$. If
there exists a connection $\G$ on $\Y$ such that the 
diagram
\be
\begin{array}{rcccl}
 & Y' & \op\longrightarrow^{\G'} & J^1Y &  \\
 _{i_Y} & \put(0,-10){\vector(0,1){20}} & & \put(0,-10){\vector(0,1){20}} &
_{J^1i_Y} \\
 & Y & \op\longrightarrow^\G & {J^1Y'} & 
 \end{array} 
 \ee
commutes, $\G'$ is said to be reducible to the connection
$\G$.

\glos{The curvature and the torsion}

Let $\G$ be a connection on a fibre bundle $Y\to X$. 
Given vector fields $\tau$, $\tau'$ on $X$ and their horizontal lifts
$\G\tau$ and $\G\tau'$ (\ref{b1.85}) on $Y$, let us compute
the vector field  
\mar{160a}\beq
R(\tau,\tau')=-\G [\tau,\tau'] + [\G \tau, \G \tau'] \label{160a} 
\eeq
on $Y$. It is readily observed that this is the
vertical vector field 
\mar{160}\ben
&& R(\tau,\tau') = \tau^\la \tau'^\m R_{\la\m}^i\dr_i, \nonumber \\
&& R_{\la\m}^i = \dr_\la\G_\m^i - \dr_\m\G_\la^i +
\G_\la^j\dr_j \G_\m^i - \G_\m^j\dr_j \G_\la^i. \label{160}
\een
The $VY$-valued
horizontal 2-form on $Y$
\mar{161'}\beq
R = \frac12 R_{\la\m}^i dx^\la\wedge dx^\m\otimes\dr_i \label{161'}
\eeq
is
called the curvature of the connection $\G$. 
In an equivalent way, the curvature (\ref{161'}) is defined as the
Nijenhuis differential
\mar{1178a}\beq
R=\frac{1}{2} d_\G\G=\frac{1}{2} [\G,\G]_{\rm FN}:Y\to \op\w^2T^*X\ox VY.
\label{1178a}
\eeq

Given a connection $\G$ and a soldering form $\si$, the 
torsion of $\G$ with respect to $\si$ is defined as 
\mar{1190}\ben
&& T = d_\G \si = d_\si \G :Y\to \op\w^2 T^*X\ox VY, \nonumber\\
&& T = (\dr_\la\si_\m^i + \G_\la^j\dr_j\si_\m^i -
\dr_j\G_\la^i\si_\m^j) dx^\la\w dx^\m\ox \dr_i. \label{1190}
\een

\glos{Linear and affine connections}

A connection $\G$ on a vector bundle $Y\to X$
is said to be a linear connection if the
section 
\mar{167}\beq
\G =dx^\la\ot(\dr_\la + \G_\la{}^i{}_j(x) y^j\dr_i) \label{167}
\eeq
of the affine jet bundle $J^1Y\to Y$ is a linear bundle mrphism
over $X$.

The curvature $R$ (\ref{161'}) of a
linear connection $\G$ (\ref{167}) reads
\mar{mos4}\ben
&&R=\frac12 R_{\la\m}{}^i{}_j(x)y^j dx^\la\x dx^\m\ot\dr_i,\nonumber\\
&& R_{\la\m}{}^i{}_j = \dr_\la \G_\m{}^i{}_j - \dr_\m
\G_\la{}^i{}_j + \G_\la{}^h{}_j \G_\m{}^i{}_h - \G_\m{}^h{}_j
\G_\la{}^i{}_h. \label{mos4}
\een
Due to the vertical splitting (\ref{1.10}), it can also be represented by the
vector-valued form 
\mar{+102}\beq
R= \frac12 R_{\la\m}{}^i{}_jy^j dx^\la\w dx^\m\ot e_i. \label{+102}
\eeq

There are the following standard operations with linear connections.

(i) Let $Y\to X$ be a vector bundle and
$\G$ a linear connection (\ref{167}) on $Y$. It defines 
the dual linear connection 
\mar{b1.91}\beq
\G^*=dx^\la\ot(\dr_\la-\G_\la{}^j{}_i y_j\dr^i) \label{b1.91}
\eeq
on the dual bundle $Y^*\to X$.

(ii) Let $Y\to X$ and $Y'\to X$ be vector bundles with linear
connections $\G$ 
and $\G'$, respectively. Then the product
connection  (\ref{b1.96}) is the direct sum connection
$\G\oplus\G'$ on the Whitney sum $Y\oplus Y'$.

(iii) Let $Y\to X$ and $Y'\to X$ be vector bundles with linear
connections
$\G$ and $\G'$, respectively. They define 
the tensor product connection
\mar{b1.92}\beq
\G\ot\G'=dx^\la\ot(\dr_\la +(\G_\la{}^i{}_j y^{jk}+\G'_\la{}^k{}_r y^{ir})
\dr_{ik})
\label{b1.92}
\eeq
on the tensor product
$Y\op\otimes_X Y'\to X$.

For instance, given a linear connection $K$ (\ref{B}) on the tangent
bundle $TX\to X$, the dual connection on the cotangent bundle
$T^*X$ is 
\mar{C}\beq
K^*= dx^\la\otimes (\dr_\la -K_\la{}^\m{}_\n\dot x_\m
\dot\dr^\n). \label{C}
\eeq
Then, using the construction of the tensor product connection
(\ref{b1.92}), one can introduce the corresponding linear connection on
an arbitrary tensor bundle $T$ (\ref{mos6}).

It should be emphasized that the expressions (\ref{B}) and
(\ref{C}) for a world connection
differ in a minus sign from those usually used in the physical literature.

The curvature of a world connection is defined as the curvature
$R$ (\ref{+102}) of the connection $K$ (\ref{B}) on the tangent bundle
$TX$. It reads
\mar{1203}\ben
&& R=\frac12R_{\la\m}{}^\al{}_\bt\dot x^\bt dx^\la\w dx^\m\ot\dot\dr_\al,
\nonumber\\
&& R_{\la\m}{}^\al{}_\bt = \dr_\la K_\m{}^\al{}_\bt - \dr_\m
K_\la{}^\al{}_\bt + K_\la{}^\g{}_\bt K_\m{}^\al{}_\g -
K_\m{}^\g{}_\bt K_\la{}^\al{}_\g. \label{1203} 
\een

By the torsion of a world connection is meant the torsion
(\ref{1190}) of the connection $\G$ (\ref{B}) on the tangent bundle
$TX$ with respect to the canonical soldering form $\th_J$
(\ref{z117'}):
\mar{191}\beq
T =\frac12
T_\m{}^\n{}_\la  dx^\la\w dx^\m\ot \dr_\n,  \qquad
T_\m{}^\n{}_\la  = \G_\m{}^\n{}_\la - \G_\la{}^\n{}_\m.
\label{191}
\eeq

\glos{Affine connections}

Let $Y\to X$ be an affine bundle modelled over a vector bundle $\ol Y\to X$.
A connection $\G$ on $Y\to X$ is said to be an
affine connection if the section $\G$
is an affine bundle morphism over $X$. 

For any affine connection $\G:Y\to J^1Y$, the corresponding
linear derivative $\ol \G:\ol Y\to J^1\ol Y$ (\ref{1355'}) defines uniquely
the associated linear connection on the vector bundle $\ol Y\to X$.
Since every vector bundle has a natural structure of an affine bundle,
any linear connection on a vector bundle is also an affine connection.

Using affine bundle coordinates $(x^\la,y^i)$ on $Y$, an affine
connection $\G$ on $Y\to X$ reads
\mar{184}\beq
\G_\la^i=\G_\la{}^i{}_j(x) y^j + \si_\la ^i(x). \label{184}
\eeq
The coordinate expression of the associated linear connection is 
\mar{mos032}\beq
\ol\G_\la^i=\G_\la{}^i{}_j(x) \ol y^j, \label{mos032}
\eeq
where $(x^\la,\ol y^i)$ are the associated linear bundle coordinates on
$\ol Y$.

Affine connections on an affine bundle $Y\to X$ constitute an affine space
modelled over the soldering forms on $Y\to X$. In view of the vertical
splitting (\ref{48}), these soldering forms can be seen as global sections of
the vector bundle $T^*X\ot\ol Y\to X$. If $Y\to X$ is a vector bundle,
both the affine connection $\G$ (\ref{184}) and the associated linear
connection $\ol\G$ are connections on the same vector bundle $Y\to X$, and
their difference is a basic soldering form on $Y$. Thus, every affine
connection on a vector bundle $Y\to X$ is the sum of a linear connection 
and a basic soldering form on $Y\to X$.

Given an affine connection $\G$ on a vector bundle
$Y\to X$, let $R$ and $\ol R$ be the curvatures of
the connection $\G$ and the associated linear connection $\ol \G$,
respectively.  It is readily observed that $R = \ol R + T$,
where the $VY$-valued 2-form
\mar{mos036}\ben
&& T=d_\G\si=d_\si\G :X\to \op\wedge^2 T^*X\op\otimes_X VY, \nonumber \\
&& T =\frac12 T_{\la
\m}^i dx^\la\wedge dx^\m\otimes \dr_i,  \qquad
T_{\la \m}^i = \dr_\la\si_\m^i - \dr_\m\si_\la^i + 
\si_\la^h \G_\m{}^i{}_h - \si_\m^h \G_\la{}^i{}_h, \label{mos036}
\een
is the torsion (\ref{1190}) of the connection $\G$
with respect to the basic soldering form $\si$. 

In particular, let us consider the tangent bundle $TX$ of a manifold
$X$. We have the canonical soldering form $\si=\th_J=\th_X$ (\ref{z117'}) on
$TX$. Given an arbitrary 
world connection $K$ (\ref{B}) on $TX$, the corresponding 
affine connection
\mar{b1.97}\beq
A=K +\th_X, \qquad
A_\la^\m=K_\la{}^\m{}_\n \dot x^\n +\dl^\m_\la, \label{b1.97}
\eeq
on $TX$ is called the Cartan connection. 
The torsion of the Cartan connection coincides with the torsion
$T$ (\ref{191}) of the  world connection $K$, while its curvature
is the sum $R+T$ of the curvature and the torsion of $K$.

\glos{Composite connections}

Let us consider the composition 
\mar{1.34}\beq
Y\to \Si\to X, \label{1.34}
\eeq
of fibre bundles
\mar{z275,6}\ben
&& \pi_{Y\Si}: Y\to\Si, \label{z275}\\
&& \pi_{\Si X}: \Si\to X. \label{z276}
\een
It is called a composite fibre bundle, and is 
provided with an atlas of bundle coordinates $(x^\la,\si^m,y^i)$, where
$(x^\m,\si^m)$ are bundle coordinates on the fibre bundle (\ref{z276}) and the
transition functions $\si^m\to\si'^m(x^\la,\si^k)$ are independent of the
coordinates $y^i$.
The following two assertions make composite fibre bundles  useful for physical
applications.
     
\begin{prop}\label{comp10} \mar{comp10}
Given a composite fibre bundle (\ref{1.34}), let $h$ be a global section
of the fibre bundle $\Si\to X$. Then the restriction
\mar{S10}\beq
Y_h=h^*Y \label{S10}
\eeq
of the fibre bundle $Y\to\Si$ to $h(X)\subset \Si$ is a subbundle 
$i_h: Y_h\hookrightarrow Y$
of the fibre bundle $Y\to X$. 
\end{prop}

\begin{prop} \label{mos61} \mar{mos61}  Given
a section $h$ of the fibre bundle
$\Si\to X$ and a section $s_\Si$ of the fibre bundle $Y\to\Si$, their
composition
\mar{1.37}\beq
s=s_\Si\circ h \label{1.37}
\eeq
is a section of the composite fibre  bundle $Y\to X$ (\ref{1.34}).
Conversely, every section $s$ of the fibre bundle $Y\to X$ is the
composition (\ref{1.37}) of the section $h=\pi_{Y\Si}\circ s$ of the
fibre bundle $\Si\to X$ and some section $s_\Si$ of the fibre
bundle $Y\to \Si$ over the closed submanifold $h(X)\subset \Si$.
\end{prop}

Let 
\mar{b1.113}\beq
 A_\Si=dx^\la\ot (\dr_\la + A_\la^i\dr_i) +d\si^m\ot (\dr_m + A_m^i\dr_i) 
\label{b1.113}
\eeq
be a connection on the fibre bundle $Y\to \Si$.
Let $h$ be a section of the fibre bundle $\Si\to X$
and $Y_h$ the subbundle (\ref{S10}) of the composite fibre bundle $Y\to X$,
which is the restriction of the fibre bundle $Y\to\Si$ to $h(X)$. 
Every connection $A_\Si$ (\ref{b1.113}) induces the pull-back
connection
\mar{mos83}\beq
A_h=i_h^*A_\Si=dx^\la\ot[\dr_\la+((A^i_m\circ h)\dr_\la h^m
+(A\circ h)^i_\la)\dr_i] \label{mos83}
\eeq
on $Y_h\to X$.

Given a composite fibre bundle $Y$ (\ref{1.34}), there is the following exact
sequences of vector bundles over $Y$:
\mar{63a}\beq
0\to V_\Si Y\hookrightarrow VY\to Y\op\times_\Si V\Si\to 0, \label{63a}
\eeq
where $V_\Si Y$ is vertical tangent bundle of the fibre bundle
$Y\to\Si$.
Every connection $A$ (\ref{b1.113}) on the fibre bundle $Y\to\Si$ provides
the splitting
\mar{46a}\ben
&& VY=V_\Si Y\op\oplus_Y A_\Si(Y\op\times_\Si V\Si),\label{46a}\\
&& \dot y^i\dr_i + \dot\si^m\dr_m=
(\dot y^i -A^i_m\dot\si^m)\dr_i + \dot\si^m(\dr_m+A^i_m\dr_i), \nonumber 
\een
of the exact sequence (\ref{63a}).
Using this splitting (\ref{46a}), one can construct the first order
differential operator  
\mar{7.10}\beq
\wt D: J^1Y\to T^*X\op\otimes_Y V_\Si Y,
\qquad
\wt D= dx^\la\otimes(y^i_\la-
A^i_\la -A^i_m\si^m_\la)\dr_i, \label{7.10}
\eeq
called the vertical covariant differential, 
on the composite fibre bundle $Y\to X$. It possesses the following
important property. Let $h$ be a section of the fibre bundle $\Si\to X$
and $Y_h$ the subbundle (\ref{S10}) of the composite fibre bundle $Y\to X$,
which is the restriction of the fibre bundle $Y\to\Si$ to $h(X)$. Then the
restriction of the vertical covariant differential $\wt D$
(\ref{7.10}) to $J^1i_h(J^1Y_h)\subset J^1Y$ coincides with the
familiar covariant differential on $Y_h$ relative to the pull-back
connection $A_h$ (\ref{mos83}).

\glos{Principal connections}

Let $\pi_P :P\to X$ be a principal bundle whose structure group is
a real Lie group $G$. 
By definition, $P\to X$ is provided with 
the free transitive action of $G$ on $P$ on the
right:
\mar{1}\beq
R_G:P\op\times_X G \to P, \qquad
R_g : p\mto pg, \quad p\in P,\quad g\in G. \label{1}
\eeq
A $G$-principal bundle $P$ is equipped with a bundle atlas
$\Psi_P=\{(U_\al,\psi^P_\al),\rho_{\al\bt}\}$ whose trivialization morphisms 
\be
\s_\al^P :\p_P^{-1}(U_\al)\to
U_\al\times G 
\ee
obey the condition
\be
\pr_2\circ \s_\al^P\circ R_g=g\circ \pr_2\circ \s_\al^P, \qquad \forall g\in G.
\ee
Due to this property, every trivialization morphism $\psi^P_\al$
determines a unique local section $z_\al:U_\al\to P$ such that
$\pr_2\circ \psi^P_\al\circ z_\al=\id$. The
transformation rules for $z_\al$ read
\mar{b1.202}\beq
z_\bt(x)=z_\al(x)\rho_{\al\bt}(x),\qquad x\in U_\al\cap U_\bt.\label{b1.202}
\eeq
Conversely, the family $\{(U_\al,z_\al)\}$ of local sections 
of
$P$ which obey (\ref{b1.202}) uniquely determines a bundle atlas
$\Psi_P$ of $P$.

The pull-back $f^*P$ (\ref{mos106}) of a principal bundle is also a
principal bundle with the same structure group. 

Taking the quotient of the tangent bundle $TP\to P$ and the vertical tangent
bundle $VP\to P$ of $P$ by $G$, we obtain the
vector bundles
\mar{b1.205}\beq
 T_GP=TP/G,\qquad   V_GP=VP/G \label{b1.205}
\eeq
over $X$. Sections of $T_GP\to X$ are
$G$-invariant vector fields on $P$, while sections of
$V_GP\to X$ are
$G$-invariant vertical vector fields on $P$. 
Hence, the typical fibre
of $V_GP\to X$ is the right
Lie algebra ${\got g}_r$ of the group
$G$. The group $G$ acts on this typical fibre by the adjoint representation.

Let $J^1P$ be the first order jet manifold of a $G$-principal bundle $P\to
X$. 
Bearing in mind that $J^1P\to P$ is an affine bundle modelled
over the vector bundle 
\be
 T^*X\op\ot_P VP\to P,
\ee
let us consider the quotient of the jet bundle $J^1P\to P$ by
the jet prolongation of the canonical action (\ref{1}). We
obtain the affine bundle  
\mar{B1}\beq
C=J^1P/G\to X\label{B1}
\eeq
 modelled over the vector bundle 
\be
\ol C=T^*X\ox V_GP\to X. 
\ee

Turn now to connections on a principal bundle $P\to X$. In this case,
the exact sequence (\ref{1.8a}) can be reduced to the exact sequence
\mar{1.33}\beq
0\to V_GP\op\hookrightarrow_X T_GP\to TX\to 0 \label{1.33}
\eeq
by taking the quotient with respect to the action of the group $G$.
A principal connection $A$ on a principal
bundle $P\to X$ is defined as a section $A: P\to J^1P$ which is
equivariant under the action (\ref{1}) of the group $G$ on $P$, i.e.,
\mar{b1.210}\beq
J^1R_g\circ A= A\circ R_g, \qquad \forall g\in G. \label{b1.210}
\eeq
Such 
a connection defines the splitting of the exact sequence (\ref{1.33}),
and can be represented by the $T_GP$-valued form
\mar{1131}\beq
A=dx^\la\ot (\dr_\la + A_\la^q \ve_q),  \label{1131}
\eeq
where $\{\ve_q\}$ is the basis for the Lie algebra ${\got g}_r$.

On the other hand, due to the property (\ref{b1.210}), there is 
obvious one-to-one correspondence between the principal connection
on a principal bundle $P\to X$ and the global sections of the
fibre bundle $C\to X$ (\ref{B1}), called the bundle of
principal connections. Given a bundle atlas of $P$, the fibre bundle 
$C$ is equipped with the associated bundle
coordinates $(x^\la,a^q_\la)$ such that, for any section $A$ of $C\to
X$, the local functions $A^q_\la=a^q_\la\circ A$ are coefficients of
the connection form (\ref{1131}). One can show that they coincide with
coefficients of the familiar local connection form \cite{kob} and,
therefore,  can
be treated as gauge potentials in gauge theory on a $G$-principal bundle $P$.

There are both pull-back and push-forward
operations of principal connections \cite{kob}.

\begin{theo} \label{mos252} \mar{mos252}
Let $P$ be a principal fibre bundle and $f^*P$ (\ref{mos106})
the pull-back principal bundle with the same structure group.
If
$A$ is a principal connection on $P$, then the pull-back connection $f^*A$
(\ref{mos82}) on $f^*P$ is a principal connection.
\end{theo}

\begin{theo} \label{mos253} \mar{mos253}
Let $P'\to X$ and $P\to X$ be principle bundles with structure groups
$G'$ and $G$, respectively. Let $\Phi: P'\to P$ be a principal bundle morphism
over
$X$ with the corresponding homomorphism $G'\to G$. For every principal
connection
$A'$ on $P'$, there exists a unique principal connection
$A$ on $P$  such that $T\Phi$ sends the
horizontal subspaces of $A'$ onto the horizontal subspaces of $A$. 
\end{theo}

Let the structure group $G$ of a principal bundle $P$
acts on some manifold $V$ on the left. Let us consider the quotient
\mar{b1.230}\beq
Y=(P\times V)/G \label{b1.230}
\eeq
by identification of the elements
$(p,v)$ and $(pg,g^{-1}v)$ for all $g\in G$. It is a fibre bundle over
$X$ called a $P$-associated fibre bundle.
Every atlas $\Psi_P=\{(U_\al,z_\al)\}$ of
$P$ determines 
the associated atlas 
$\Psi=\{(U_\al,\psi_\al(x)=[z_\al(x)]^{-1})$ of $Y$.
Any automorphism $\Phi$ of a principal bundle $P$ yields the 
automorphism 
\be
\Phi_Y:(P\times V)/G\to (\Phi(P)\times V)/G
\ee
of the $P$-associated fibre bundle (\ref{b1.230}).

Every principal connection on $P\to X$ induces canonically the
corresponding connection on the $P$-associated fibre bundle (\ref{b1.230})
as follows. Given a principal connection $A$ (\ref{1131}) on $P$
and the corresponding horizontal splitting of the tangent bundle $TP$,
the tangent map to the canonical morphism 
\be
P\times V\to (P\times V)/G
\ee
 defines the
horizontal splitting of the tangent bundle $TY$ and the
corresponding connection on the $P$-associated fibre bundle $Y\to X$
\cite{kob}. The latter is called the associated principal
connection or simply a principal
connection on $Y\to X$.
If $Y$ is a vector bundle, this connection takes the form
\be
A=dx^\la\ot(\dr_\la -A^p_\la I^i_p\dr_i), 
\ee
where $I_p$ are generators of the representation of the Lie
algebra ${\got g}_r$ in $V$.

\end{document}